\documentclass[a4paper,hyper]{JHEP3}
\usepackage[pdftex]{graphicx}
\newcommand{\shalf}{\ensuremath{\textstyle \frac{1}{2}}}

\newcommand{\kdag}{\mathbin{k\mkern-10mu\big/}}

\newcommand{\pp}{{\scriptscriptstyle ||}}

\def\lsim{\mathrel{\raise.3ex\hbox{$<$\kern-.75em\lower1ex\hbox{$\sim$}}}}
\def\gsim{\mathrel{\raise.3ex\hbox{$>$\kern-.75em\lower1ex\hbox{$\sim$}}}}

\def\Slashnew#1{#1\kern-0.55em\raise.05ex\hbox{/}}
\def\slashnew#1{#1\kern-0.5em\raise.05ex\hbox{{$\scriptstyle /$}}}
\def\sfrac#1#2{{\textstyle\frac#1#2}}

\def\prop{\Delta}

\def\ie{{ i.e. }}
\def\eg{{ e.g. }}

\newcommand{\beq} {\begin{equation}}
\newcommand{\eeq} {\end{equation}}
\newcommand{\beqa}{\begin{eqnarray}}
\newcommand{\eeqa}{\end{eqnarray}}

\title{Kinetic theory for scalar fields with nonlocal quantum coherence}

\author{Matti Herranen,
        Kimmo Kainulainen and
        Pyry Matti Rahkila 
        \\
        \\ Department of Physics, P.O.Box 35 (YFL), 
        \\ FIN-40014 University of Jyv\"askyl\"a, Finland, 
        \\ and 
  	    \\ Helsinki Institute of Physics, P.O.~Box 64, 
  	    \\ FIN-00014 University of Helsinki, Finland.\\
        \\e-mail: \email{matherr@phys.jyu.fi,
                         kainulai@phys.jyu.fi
                         pmrahkil@phys.jyu.fi}}

\abstract{We derive quantum kinetic equations for scalar fields undergoing
coherent evolution either in time (coherent particle production) or in
space (quantum reflection). Our central finding is
that in systems with certain space-time symmetries, quantum coherence
manifests itself in the form of new spectral solutions for the
dynamical 2-point correlation function. This spectral structure leads
to a consistent approximation for dynamical equations that
describe coherent evolution in presence of decohering collisions. We
illustrate the method by solving the bosonic Klein problem and the bound 
states for the nonrelativistic square well potential. We then compare our 
spectral phase space definition of particle number to other definitions in 
the nonequilibrium field theory. Finally we will explicitly compute the
effects of interactions to coherent particle production in the case of
an unstable field coupled to an oscillating background.}

\preprint{}

\begin{document}
\section{Introduction}
Quantum transport effects are gaining more and more interest in many
applications in modern particle physics and cosmology. This is true in
particular for the case of the electroweak
baryogenesis~\cite{EWBGrew,CJK,Earlier,KPSW,PSW}, but also for the
leptogenesis~\cite{buchmuller} or particle creation in the early
universe~\cite{Prokopec_partnumber} and during phase
transitions~\cite{brandenbeger}. We have recently developed new
quantum transport equations for fermionic systems including nonlocal
coherence, either in space (quantum reflection) or in time (coherent
pair production)~\cite{HKR1,HKR2}, including the effects of decohering
collisions~\cite{HKR2}. Here we will introduce a similar formalism for
the scalar fields. As in the fermionic case, the coherence information
is found to be encoded in new spectral solutions in the phase space of
the dynamical 2-point function. The physical information of particle
numbers or fluxes and of coherence is carried by a set of scalar
functions that parametrize the spectral shells in the full 2-point
correlation function.

Our approach can be summarized as follows. We first formulate
Schwinger-Dyson equations for the 2-point correlation functions using
the Closed Time Path (CTP) method. We solve the resulting
  Kadanoff-Baym (KB) equations for the
2-point Wightman function $\prop^<$ in the noninteracting mean field
limit in the mixed representation. The most general solution for
$\prop^<$ in this limit is a sum of singular spectral distribution
functions corresponding to the usual mass-shell solutions with a
dispersion relation $\omega^2 = \vec k^2 + m^2$, and a new coherence
solutions living at shell $k_z= 0$ in the static planar symmetric case
or at shell $k_0=0$ in the case of a spatially homogeneous
system. We then turn back to the full KB-equations including collision
terms and integrate them with an appropriate set of moments. 
On the adiabatic boundaries the lowest moments of $\prop^<$ can be 
related to the spectral on-shell functions. 
From practical point of view the most important aspect of
our method is that the singular shell structure reduces integrated 
equations of motion, including the collision terms,
to a closed set of equations for spectral on-shell functions, or 
equivalently for a finite number of lowest moments of
$\prop^<$ (three in case of a single real scalar field).

Since the singular shell representation of the coherence is so crucial
for our formalism, we have given several examples which illustrate
their physical role. Firstly, we will solve the bosonic Klein
problem. We will show that in the absence of the coherence shell the
quantum nature of reflection is completely lost, but that the correct
reflection and tunneling factors are recovered when the coherence
shell is included. We will also show that the spectral phase
space definition of particle number in our formalism is consistent
with other definitions for nonequilibrium systems in the
literature~\cite{Berges}. In particular, our particle number, when
applied to Bunch-Davies vacuum in the inflation, corresponds to the
adiabatic particle number that remains always zero in a conformally
coupled scalar theory~\cite{BirrelAndDavis}. The Klein problem example
also allows us to demonstrate how the on-shell functions are related
to moment functions that must be used to formulate a dynamical problem
with only an incomplete information about the variables defining the
system. We will also consider production of unstable scalar
particles by a coherent time dependent background potential and
decoherence and thermalization of an initially highly correlated
state.

This paper is organized as follows. In section \ref{sec:formalism} we
briefly review the basic CTP-formulation for the calculation of the
2-point function and in section~\ref{sec:shell} we derive the spectral
shell structure of the Wightman function in the mean field limit.  In
section \ref{sec:physical} we use our formalism to solve the bosonic
Klein problem. We also derive an expression for our on-shell particle
number in terms of moment functions and compare it with other
definitions in the literature and apply it to the particle production
during inflation. We also compute other measurable quantities
such as energy density and the pressure. In section
\ref{sec:non-relativistic} we solve the nonrelativistic problem with a
Schr\"odinger equation and show that there one obtains similar
coherence solutions for the description of reflection in the planar
symmetric case. We complete this section by solving the bound states of the square well
  potential with our formalism. In section \ref{sec:spectral} we show
  that, just as with fermions, the coherence solutions are excluded
  from the spectral function by the spectral sum rule. In section
  \ref{sec:interactions} we derive the dynamical (moment) equations
  for a scalar field including collisions for a spatially homogeneous
  system. In section \ref{sec:numexample} we consider coherent production of
  unstable particles. Finally section
  \ref{sec:discussion} contains our discussion and outlook.
\section{CTP-formalism for scalars}
\label{sec:formalism}

The main object of interest for us is the 2-point {\em Wightman}
function $\prop^<$ for a real scalar field, defined as:
\begin{equation}
  i\prop^<(u,v) \equiv
         \langle  \phi(v)\phi(u)\rangle \equiv
         {\rm Tr}\left\{\hat \rho \ \phi(v)\phi(u)\right\},
\label{G-less}
\end{equation}
where $\hat \rho$ is some unknown quantum density operator that gives
the complete information of the system. Instead of trying to find a
solution for $\hat \rho$, we will set up equations for the ``in-in"
correlation function (\ref{G-less}) using the Schwinger-Keldysh or
Closed Time Path (CTP) formalism~\cite{Schwinger-Keldysh,CTP}. In this
formalism one first defines 2-point correlation function on a
complex time-path:
\begin{equation}
  i\prop_{\cal C}(u,v) =
            \left\langle T_{\cal C}
                 \left[\phi(u) \phi (v)\right]
             \right\rangle,
\label{Gcontour}
\end{equation}
where $T_{\cal C}$ defines time ordering along the Keldysh contour
shown in figure~\ref{fig:KeldyshPath}.
\FIGURE[t]{\includegraphics[width=9cm]{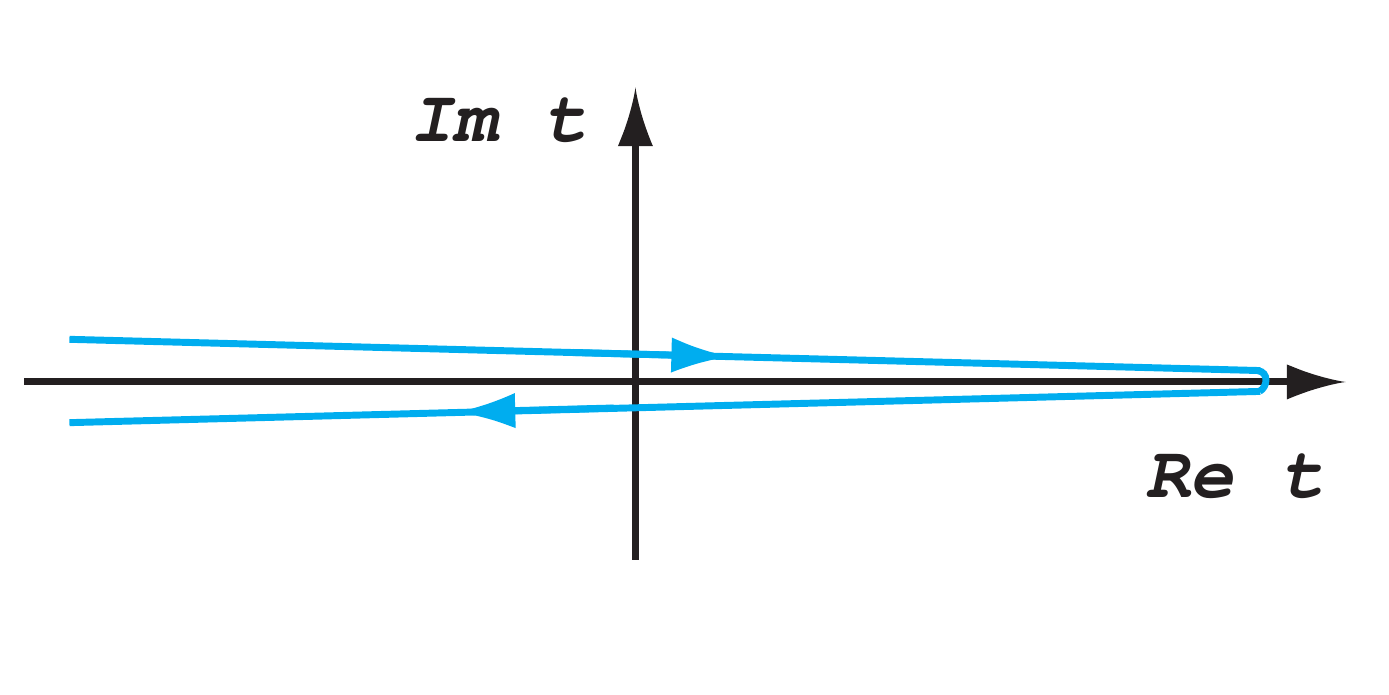}
        \caption{Schwinger-Keldysh path in complex time.}
        \label{fig:KeldyshPath}}
One can show, for example by use of the two-particle-irreducible (2PI)
effective action techniques \cite{CTP,2PI}, that $\prop_{\cal C}(x,y)$
obeys the contour Schwinger-Dyson equation:
\begin{equation}
\prop_{\cal C} (u,v) = \prop^0_{\cal C} (u,v)
             + \int_{\cal C} {\rm d}^4z_1 \int_{\cal C}
                             {\rm d}^4z_2 \; \prop^0_{\cal C} (u,z_1)
               \Pi_{\cal C} (z_1,z_2) \prop_{\cal C} (z_2,v)\,,
\label{SD1}
\end{equation}
where $\prop^0_{\cal C}$ is the free propagator of the theory, and the precise form of the self-energy function $\Pi_{\cal C}$ depends on the Lagrangian and the truncation scheme. Once the theory is specified, it can be computed from the 2PI-effective action by functional differentiation:
\begin{equation}
\Pi_{\cal C}(u,v) \equiv i\frac{\delta \Gamma_2[G]}{\delta
  \prop_{\cal C}(v,u)} \,,
\label{2PIsigmasC}
\end{equation}
where $\Gamma_2$ is the sum of all two particle irreducible vacuum graphs in the theory. The complex time Green's function in (\ref{Gcontour}) can be  decomposed in four different 2-point functions with respect to usual real time variable:  
\begin{eqnarray}
  i\prop^<(u,v) \equiv i\prop^{+-}(u,v) &\equiv& 
\langle \phi(v)\phi(u) \rangle 
   \nonumber\\
  i\prop^>(u,v) \equiv i\prop^{-+}(u,v) &\equiv& 
\langle \phi(u)\phi(v) \rangle
   \nonumber\\
  i\prop_F(u,v) \equiv i\prop^{++}(u,v) &\equiv&   
\theta(u_0-v_0) i\prop^>(u,v) + \theta(v_0-u_0) i\prop^<(u,v)
   \nonumber\\
  i\prop_{\bar F}(u,v) \equiv i\prop^{--}(u,v) &\equiv&
  \theta(v_0-u_0) i\prop^>(u,v) + \theta(u_0-v_0)i\prop^<(u,v)\,,
\label{GFs}
\end{eqnarray}
where $\prop_F$ and $\prop_{\bar F}$ are the chronological (Feynman)
and anti-chronological (anti-Feynman) Green's functions, respectively,
and $\prop^<$ and $\prop^>$ are the Wightman functions we are
primarily interested in solving here. A similar decomposition can be
done for the contour self-energy $\Pi_{\cal C}$ to get:
\begin{equation}
\Pi^{ab}(u,v) \equiv iab \frac{\delta \Gamma_2[G]}{\delta
  \prop^{ba}(v,u)} \,,
\label{2PIsigmas}
\end{equation}
where the indices $a,b$ refer to the position of the arguments $u$ and $v$, respectively, on the complex Keldysh time path. When $a=+1(-1)$ the time argument in $u$ belongs to the upper (lower) branch in figure \ref{fig:KeldyshPath}, and we will use the same notation: $\Pi^< = \Pi^{+-}$, etc.~for the self energy as we did for the propagators (\ref{GFs}). It can then be shown that the complex-time equation (\ref{SD1}) is equivalent to the following matrix equation with a usual real time argument: 
\begin{equation}
\prop_0^{-1} \otimes \prop = \sigma_3 \; \delta +  \Pi \otimes \sigma_3 \prop,
\label{SD3}
\end{equation}
where
\begin{equation}
\prop=\left(\begin{array}{cc}
            \prop_F & \prop^< \\
            \prop^> & \phantom{-} \prop_{\bar F}
         \end{array}\right) \qquad , \qquad 
\Pi=\left(\begin{array}{cc}
            \Pi_F & \Pi^< \\
            \Pi^> & \phantom{-} \Pi_{\bar F} \,,
         \end{array}\right) 
\label{Gmatrix}
\end{equation}
and $\sigma_3$ is the usual Pauli matrix, and we defined a shorthand
notation $\otimes$ for the convolution integral:
\begin{equation}
    f \otimes g \equiv \int {\rm d}^4z f(u,z)g(z,v).
\label{otimes}
\end{equation}
We have also left out the labels $u$ and $v$ where obvious; for example
$\delta \equiv \delta^4(u-v)$.
\subsection{Kadanoff-Baym equations}

It's appropriate to further define the retarded and advanced
propagators (a similar decomposition obviously holds for the self energy  
function $\Pi$):
\begin{eqnarray}
  \prop^r(u,v)  &\equiv&  \phantom{-} \theta(u^0-v^0) (\prop^> -
  \prop^<) \nonumber\\ 
  \prop^a(u,v)  &\equiv&  -\theta(v^0-u^0) (\prop^> - \prop^<).
\label{raGFs}
\end{eqnarray}
Moreover, the hermiticity properties of the Wightman functions: 
\begin{equation}
  \left[i\prop^{<,>}(u,v)\right]^\dagger = i\prop^{<,>}(v,u)
\label{CEq1}
\end{equation}
imply that $\left[i\prop^{r}(u,v)\right]^\dagger = -i\prop^{a}(v,u)$, which suggests a decomposition into hermitian and antihermitian parts:  
\begin{eqnarray}
  \prop_H       &\equiv&  \frac{1}{2}\left(\prop^a + \prop^r\right) \nonumber\\ 
  {\cal A}  &\equiv&  \frac{1}{2i}\left(\prop^a - \prop^r\right) 
                     = \frac{i}{2}\left(\prop^> - \prop^<\right).  
\label{Hdecomposition}
\end{eqnarray}
The antihermitian part ${\cal A}$ is called the \textit{spectral function}. Based on (\ref{raGFs}) it is easy to show that $\prop_H$ and ${\cal A}$ obey the spectral relation: $\prop_H(u,v) = -i {\rm sgn}(u^0-v^0) {\cal A} (u,v).$ Since the self-energies $\Pi$ satisfy identities similar to (\ref{CEq1}), we can define the hermitian and antihermitian parts of $\Pi^{r,a}$ as well:
\begin{eqnarray}
  \Pi_H  &\equiv&  \frac{1}{2}\left(\Pi^a + \Pi^r\right) \nonumber\\ 
  \Gamma    &\equiv&  \frac{1}{2i}\left(\Pi^a - \Pi^r\right) 
                     = \frac{i}{2}\left(\Pi^> - \Pi^<\right)\,.  
\label{gamma}
\end{eqnarray}
Using the definitions (\ref{Hdecomposition})-(\ref{gamma}) it is now
straightforward to show that eqs.~(\ref{SD3}), when written in the component form, become: 
\begin{eqnarray}
   (\prop_0^{-1}-\Pi_H) \otimes {\cal A} - \Gamma
   \otimes \prop_H = 0
\nonumber \\
    (\prop_0^{-1}-\Pi_H) \otimes \prop_H + \Gamma
   \otimes {\cal A} = \delta 
\label{SpecEq1}
\end{eqnarray}
and
\begin{equation}
   (\prop_0^{-1}-\Pi_H) \otimes \prop^< - \Pi^< \otimes \prop_H
   = \frac{1}{2}\left( \Pi^> \otimes \prop^< - \Pi^< \otimes \prop^>  
\right) \,,
\label{Dyneq}
\end{equation}
where $\prop_0^{-1}$ is the inverse free propagator. Equations (\ref{SpecEq1}) are called the \textit{pole equations} and eq.~(\ref{Dyneq}) the \textit{Kadanoff-Baym (KB) equation}. The other KB-equation for $\prop^>$ need not be considered, since form the definition (\ref{Hdecomposition}) it immediately follows that $\prop^> = \prop^< - 2i {\cal A}$.

\subsection{Mixed representation}
The final step in our formal derivation is moving to the mixed representation, to separate the external and internal degrees of freedom in the correlators through a Wigner transform:
\begin{equation}
F(k,x) \equiv \int d^{\,4} r \, e^{ik\cdot r} F(x + r/2,x-r/2) \,,
\label{wigner1}
\end{equation}
where $x\equiv (u+v)/2$ is the average coordinate, and $k$ is the internal momentum variable conjugate to relative coordinate $r\equiv v-u$. Performing the Wigner transformation to eqs.~(\ref{SpecEq1}) and (\ref{Dyneq}) we get the pole-equations
\begin{eqnarray}
  \prop_0^{-1} {\cal A}
  -  e^{-i\Diamond}\{ \Pi_H \}\{ {\cal A}\}
  -  e^{-i\Diamond}\{ \Gamma \}\{\prop_H\} &=& 0
\label{SpecEqMix1} \\
  \prop_0^{-1} \prop_H
  -  e^{-i\Diamond}\{ \Pi_H \}\{ \prop_H \}
  +  e^{-i\Diamond}\{ \Gamma \}\{{\cal A}\} &=& 1
\label{SpecEqMix2}
\end{eqnarray}
and the KB-equation for $\prop^<$ becomes
\begin{equation}
 \prop_0^{-1}\prop^<
  -  e^{-i\Diamond}\{ \Pi_H \}\{ \prop^< \}
  -  e^{-i\Diamond}\{ \Pi^< \}\{ \prop_H \}
  = {\cal C}_{\rm coll} \,,
\label{DynEqMix}
\end{equation}
The collision term in eq.~(\ref{DynEqMix}) is given by
\begin{equation}
{\cal C}_{\rm coll} = -i e^{-i\Diamond}
                             \left( \{\Gamma\}\{\prop^<\} -
                                    \{\Pi^<\}\{{\cal A}\}\right)\,,
\label{collintegral}
\end{equation}
and the $\Diamond$-operator is the following generalization of the Poisson brackets:
\begin{equation}
\Diamond\{f\}\{g\} = \frac{1}{2}\left[
                   \partial_X f \cdot \partial_k g
                 - \partial_k f \cdot \partial_X g \right]\,.
\label{diamond}
\end{equation}
Equations (\ref{SpecEqMix1})-(\ref{DynEqMix}) are the master equations appropriate for all analysis to be performed in this paper. Explicit forms of $\prop_0$ and the interactions depend on the model. In this paper we consider a theory defined by the Lagrangian
\begin{equation}
{\cal L} = \frac12 (\partial_\mu \phi)^2 -\frac12 m^2 \phi^2 + {\cal L}_{\rm int} \,,
\label{freeLag1}
\end{equation}
where $m=m(x)$ is possibly spacetime dependent mass and ${\cal L}_{\rm int}$ is the interaction part to be defined later.  The inverse free propagator corresponding to eq.~(\ref{freeLag1}) in the mixed representation is
\begin{equation}
\prop_0^{-1} \equiv 
      k^2 - \frac14 \partial^2 + ik\cdot\partial 
      - m^2 e^{-\frac{i}{2}\partial^m_x \cdot \partial^\prop_k} \,,
\label{freeprop}
\end{equation}
where the $\partial^m_x$-derivative always acts on the mass term, and the $\partial^\prop_k$-derivative to the Green's function ${\cal A}$, $\Delta_H$ or $\prop^<$ in eqs.~(\ref{SpecEqMix1})-(\ref{DynEqMix}) respectively. 

\section{Shell structure}
\label{sec:shell}
In analogy to what was found in the case of fermions~\cite{HKR1,HKR2}, a reasonable approximation scheme can be developed for an interacting theory employing the spectral shell structure of the noninteracting theory. So let us first consider noninteracting fields, for which $\Pi^{ab} = 0$. In this case the KB-equation (\ref{DynEqMix}) for $\prop^<$ decouples from the pole-functions $\prop_H$ and ${\cal A}$ and reduces to Klein-Gordon equation in momentum space:
\begin{equation}
 \Big(k^2 - \frac14 \partial^2 + ik\cdot\partial - m^2 e^{-\frac{i}{2}
      \partial^m_x \cdot \partial^\prop_k} \Big)i\prop^< = 0\,. 
\label{KG_Eq1}
\end{equation}
This is still a very complicated equation because it contains derivative operators to arbitrarily high orders. We shall analyze it in more detail in the {\em mean field limit}, where $m^2
e^{-\frac{i}{2}{\partial^m_x \cdot \partial^\prop_k}} \rightarrow m^2$, assuming also particular space-time symmetries: a case with a spatial homogeneity and a static problem with a planar symmetry.

\subsection{Spatially homogeneous case}

In the spatially homogeneous case all spatial gradients of the mass
$m=m(t)$ and the correlator $\prop^<$ vanish. Breaking the equation
(\ref{KG_Eq1}) into real and imaginary parts and expanding to zeroth
order in the time derivatives acting on the mass then gives:
\begin{eqnarray}
 \Big(k_0^2 -\vec{k}^2 - m^2(t) - \frac14 \partial_t^2 \Big)i\prop^<(k,t)
      &=& 0
\label{KG_Eq_HOM2}\\
 k_0\partial_t i\prop^<(k,t) &=& 0\,.
\label{KG_Eq_HOM3}
\end{eqnarray}
At this point it is relevant to make a comparison with the similar
problem involving fermions~\cite{HKR1,HKR2}. In the fermionic case the
noninteracting KB-equations can be divided into algebraic {\em constraint} equations which define the phase space structure of the theory and to {\em dynamic} equations containing all time derivatives. Here such a division is not possible, as both equations contain derivative terms, and the shell structure is less obvious than with fermions. However, if we assume that $k_0 \neq 0$ then eq.~(\ref{KG_Eq_HOM3}) requires that $\partial_t i\prop^< = 0$ at all times and so one must also have $\partial_t^2 i\prop^< = 0$. Substituting this back to eq.~(\ref{KG_Eq_HOM2}) does lead to an algebraic equation
\begin{equation}
\left(k_0^2 -\vec{k}^2 -m^2(t)\right)i\prop^<_{\rm m-s} = 0\,.
\label{alkepraeq1}
\end{equation}
This equation has the spectral solution parametrized by $t$:
\begin{equation}
i\prop^<_{\rm m-s}(k_0,|\vec{k}|,t) = 
    2\pi\,{\rm sgn}(k_0)f_{s_{k_0}}(|\vec{k}|,t)
    \delta\big(k_0^2 -\vec{k}^2 -m^2(t)\big)\,, 
\label{SpecSolHOM}
\end{equation}
where $s_{k_0}\equiv{\rm sgn}(k_0)$. This is just the usual mass-shell dispersion relation 
\begin{equation} 
k_0 = \pm \omega_{\vec k}(t) \equiv \pm \sqrt{|\vec k|^2 + m^2(t)} \,.
\end{equation}
Of course this derivation was not exact, and the solution (\ref{SpecSolHOM}) satisfies exactly neither (\ref{KG_Eq_HOM2}) nor (\ref{KG_Eq_HOM3}), except for constant $m$ and $f_{s_{k_0}}$. The point is that equations (\ref{KG_Eq_HOM2}) and (\ref{KG_Eq_HOM3}) are actually inconsistent for a nonconstant mass, but the corrections that would bring the consistency back are proportional to  mass-gradients. The result (\ref{SpecSolHOM}) is thus correct to the lowest order in mass-gradients. In particular the effect of the second order derivative term in (\ref{KG_Eq_HOM2}) to the mass-shell structure is beyond the mean field approximation.

However, if we first set $k_0=0$, then equation (\ref{KG_Eq_HOM3}) is identically satisfied and we cannot constrain the size of the derivative terms as was done above. Instead, eq.~(\ref{KG_Eq_HOM2}) now becomes:
\begin{equation}
\partial_t^2 \bar\prop^< = - 4 \omega_{\vec k}^2(t) \bar\prop^< \,.
\label{q-cohHOM}
\end{equation}
For a constant mass the solution for this equation is
\begin{equation}
i\bar\prop^<_{\rm const}(k_0,|\vec{k}|,t) =
2\pi\bar A_{\vec k}\cos(2\omega_{\vec k} t + \delta_{\vec k}) \, \delta(k_0)\,,
\label{konstaratkaisu}
\end{equation}
where $\bar A_{\vec k}$ and $\delta_{\vec k}$ are some real constants and the $\delta$-function is explicitly taking care of the restriction to the shell $k_0=0$. For a generic time-varying mass an analytical solution for $i\bar\prop^<$ might not be available, but we can write the corresponding solution for $k_0=0$ in the spectral form:
\begin{equation}
i\bar\prop^<(k_0,|\vec{k}|,t) = 2\pi\,f_c(|\vec{k}|,t)\delta(k_0)\,,
\label{k0zerospec}
\end{equation}
where $f_c(|\vec{k}|,t)$ is some real-valued function. This
establishes that there exists a new solution living at shell
$k_0=0$. However, one can ask if and how this new solution can affect
the dynamics of the mass-shell functions $f_\pm$?  Indeed, in the
constant mass case, where $f_c$ is given by eq.~(\ref{konstaratkaisu})
and $f_{\pm}={\rm const.}$, the answer is no, as expected. To get
these solutions however, we implicitly introduced prior information on
$k_0$ that allowed a reduction to one particular shell at a time. More
generally, one might be interested in systems where only an imprecise
or even no prior information is available on $k_0$. In such cases some
integration procedure over $k_0$ must be introduced to define
observable physical quantities, and these quantities typically involve
contributions from several shells. When such integration procedure is
imposed on eqs.~(\ref{KG_Eq_HOM2})-(\ref{KG_Eq_HOM3}) they generally
lead to nontrivial mixing involving the functions $f_c$. We will
return to this procedure in more detail in section
\ref{sec:physical}. The basic issue however is that the phase space of
the free dynamical function in the noninteracting system is singular
in the mean field limit, and it contains new spectral shell at $k_0=0$
such that the most complete solution for a given momentum $|\vec{k}|$
is the combination of the solutions (\ref{SpecSolHOM}) and
(\ref{k0zerospec}): 
\begin{equation}
i\prop^<(k_0,|\vec{k}|,t) 
= i\prop^<_{\rm m-s}(k_0,|\vec{k}|,t) + i\bar\prop^<(k_0,|\vec{k}|,t)\,.
\label{fullspec_HOM}
\end{equation}
The situation is now seen to be qualitatively equivalent to the case with fermions and we interpret analogously~\cite{HKR1,HKR2} that the new $k_0=0$-solution (\ref{k0zerospec}) describes the quantum coherence between particles and antiparticles.

\subsection{Planar symmetric case}
\label{sec:planar}

Another simple geometry that allows analytic solutions is the case with $m=m(z)$ and $\partial_{t,x,y} i\prop^<= 0$, \ie a static planar
symmetric problem in the average coordinates in the Wigner 
transformation\footnote{Note that
  this actually means that the initial direct space correlator depends
  only on the internal time- or $\vec x_\pp$-separation:
  $i\prop^<(t,\vec x;t',\vec x') = i\prop^<(t-t';\vec x_\pp-\vec
  x_\pp';z,z')$. Thus all problems for which the wave equations have
  {\em stationary} solutions with $\psi \propto e^{iEt}$ (such as
  reflection problem) appear in this sense {\em static} in the Wigner
  transformed representation. Same applies of course for stationary
  dependence on $\vec x_\pp$.}.
In this case the mean field limit of the equation (\ref{KG_Eq1}) is:
\begin{eqnarray}
 \Big(k_0^2 -\vec{k}^2 - m^2 + \frac14 \partial_z^2 \Big)i\prop^<(k,z)
      &=& 0
\label{KG_Eq_STA2}\\
 k_z\partial_z i\prop^<(k,z) &=& 0\,.
\label{KG_Eq_STA3}
\end{eqnarray}
The analysis proceeds analogously to the homogeneous case. For $k_z \neq 0$, eq.~(\ref{KG_Eq_STA3}) gives $\partial_z i\prop^< = 0$ for all $z$ so that  $\partial_z^2 i\prop^< = 0$ as well, and eq.~(\ref{KG_Eq_STA2}) again reduces to the algebraic form:
\begin{equation}
\left(k_0^2 - \vec{k}^2 - m^2(z)\right)i\prop^<_{\rm m-s} = 0\,.
\label{alkepraeq2}
\end{equation}
This equation is similar to eq.~(\ref{alkepraeq1}), except that in this case energy is conserved, and the mean field momentum $k_z$ is the quantity that becomes dependent on $z$. Taking this into account we write the spectral solution in the form
\begin{equation}
i\prop^<_{\rm m-s}(k_0,|\vec{k}_\pp|,k_z,z) = 2\pi\,{\rm
  sgn}(k_0)f_{s_{k_z}}(k_0,|\vec{k}_\pp|,z)\delta\big(k_z^2 -
  k_m^2(z)\big)\,,
\label{SpecSolSTA}
\end{equation}
where
\begin{equation}
k_z = \pm k_m \equiv \pm \sqrt{k_0^2 -\vec{k}_\pp^2 - m^2(z)} \,.
\end{equation}
This is again the usual mass shell solution. We get a new solution by first setting $k_z=0$, so that eq.~(\ref{KG_Eq_STA3}) is identically satisfied and
eq.~(\ref{KG_Eq_STA2}) becomes:
\begin{equation}
 \left(\partial_z^2 + 4 k_m^2(z) \right)i\tilde\prop^< = 0\,.
\label{q-cohSTA}
\end{equation}
In constant mass limit one again finds a solution similar to eq.~(\ref{konstaratkaisu}), only now restricted to shell $k_z=0$: $i\tilde\prop^<_{\rm const} = 2\pi \tilde A\cos(2 k_m z + \tilde \delta) \,\delta(k_z)$, where $\tilde A(k_0,|\vec{k}_\pp|)$ and $\tilde \delta (k_0,|\vec{k}_\pp|)$ are some real constants. For an unspecified spatially varying mass term eq.~(\ref{q-cohSTA}) has a generic spectral solution:
\begin{equation}
i\tilde\prop^<(k_0,|\vec{k}_\pp|,k_z,z) =
2\pi\,f_c(k_0,|\vec{k}_\pp|,z)\delta(k_z)\,,
\label{kzzerospec}
\end{equation}
where $f_c(k_0,|\vec{k}_\pp|,z)$ is some real-valued function. Finally, combining the solutions (\ref{SpecSolSTA}) and (\ref{kzzerospec}) we find the most complete solution in static, planar symmetric case for given energy $k_0$ and parallel momentum $|\vec{k}_\pp|$:
\begin{equation}
i\prop^<(k_0,|\vec{k}_\pp|,k_z,z) 
= i\prop^<_{\rm m-s}(k_0,|\vec{k}_\pp|,k_z,z) +
i\tilde\prop^<(k_0,|\vec{k}_\pp|,k_z,z)\,. 
\label{fullspec_STA}
\end{equation}
In this case the mass shell solutions describe modes of left (negative direction in $z$) and right (positive direction in $z$) moving states, and the zero momentum solution their coherence.  Indeed, here the coherence solution can be pictured as quantifying the possibility that the state is simultaneously going both right and left (interference). It is therefore natural to see this solution arising at the average momentum $k_z=0$ for such a mixture.

\section{Applications}
\label{sec:physical}

Having found the spectral solutions, we wish to apply our formalism to
solve special physical problems including dynamical
evolution. Parallel to our analysis with fermions \cite{HKR1,HKR2}, we
need to define nonsingular weighted 2-point functions to replace the
singular ones (\ref{fullspec_HOM}) and (\ref{fullspec_STA}). These
functions can be given in the following generic form:
\begin{equation}
  \rho_{\cal W} (k_0,\vec{k},x)
     \equiv \int \frac{{\rm d}^4 k'}{(2\pi)^4}\; {\cal W} (k_0,
     \vec{k} \,|\, k'_0, \vec{k'}; x) \; i\prop^<(k'_0,\vec{k'},x) \,,
\label{rhoW}
\end{equation}
where the weight function ${\cal W} (k_0, \vec{k} \,|\, k'_0,
\vec{k'}; x)$ encodes our knowledge about the energy and momentum
variables of the state. (In fact a number of different weight
functions may be needed as we will see below.) Above, with the
constant mass examples, we already used implicitly weight functions
encoding precise energy or momentum resolution and below we shall give
two more examples. In the case of quantum reflection off a potential
wall (Klein problem), one may have relatively precise information on
energy, but only partial, spatially dependent information on the
momentum. In case of particle production by a homogeneous coherent
background field the momentum may be assumed to be known, but one has
no prior information on the energy, which in this case is not
conserved. 

\subsection{Klein problem}
\label{subsect:Klein}

As our first example we consider the Klein problem for scalars {\em
i.e.} a scalar particle reflecting off a planar symmetric step
potential, see figure~\ref{fig:KleinWall}. We discussed the Klein
problem in the case of fermions in~\cite{HKR1}, but the bosonic case
has some additional special characteristics. This problem could of
course be solved by use of the Klein-Gordon wave equation, but it
provides a nice setting to illustrate how to solve a dynamical problem
in our methods, which allows us to demonstrate the physicality of our
coherence shell solutions. In this case the mass is constant, but
interactions with the background potential can be represented by a
singular self-energy correction $\Pi_H$. It is easy to see that the
effect of this correction is equivalent to replacing the time
derivative with a covariant derivative: $i\partial_{v_0} \rightarrow
i \partial_{v_0} - V(v_z)$ in the inverse propagator
eq.~(\ref{freeprop}). The free KB-equation then becomes
\begin{equation}
 \Big([k_0 - V(z)]^2 e^{-\frac{i}{2}{\partial}^V_z  \partial^\prop_{k_z}} - k_z^2 + \frac14 \partial_z^2 + ik_z\partial_z - m^2
 \Big)i\prop^< = 0\,, 
\label{Klein1}
\end{equation}
where we have accounted for the fact that the $t$- and $\vec{x}_\pp$-derivatives vanish and we have taken $\vec k_\pp=0$. Taking the real and imaginary parts of eq.~(\ref{Klein1}) we find: 
\begin{eqnarray}
 \Big(k_z^2 - \frac14 \partial_z^2 - k_m^2 \cos(\sfrac12
      \partial^V_z \partial^\prop_{k_z})\Big)i\prop^<
      &=& 0
\label{Klein2} \\
 \Big(k_z \partial_z + k_m^2 \sin(\sfrac12
      \partial^V_z \partial^\prop_{k_z})\Big)i\prop^< &=& 0\,,
\label{Klein3}
\end{eqnarray}
where $k_m(z) \equiv \sqrt{(k_0-V(z))^2-m^2}$, with $V(z) = V\theta(-z)$. To proceed further we define the $n$-th moments of the Wightman function as integrals over $k_z$:
\begin{equation}
\rho_n(k_0,z)\equiv \int \frac{{\rm d} k_z}{2\pi}\;k_z^n\,
i\prop^<(k_0,\vec{k}_\pp=0,k_z,z)\,.
\label{n-moment_Klein}
\end{equation}
These integrals are convergent for arbitrary $n$ only if $i\prop^<$ is compactly 
supported or a sufficiently rapidly decreasing distribution. This is not guaranteed in general. However, in this paper we require only that the three
lowest moments $\rho_{0,1,2}$, which are related to the free two-point correlator functions, are well defined. Of course, in the adiabatic limit where eq.~(\ref{fullspec_STA}) is valid, the spectral form of the solution guarantees the existence of all moments.
These functions correspond to the weighted 2-point functions $\rho_{\cal W}$ eq.~(\ref{rhoW}) with specific weights  ${\cal W}_n = (2\pi)^3 k_z^n \delta(k_0-k'_0) \delta^2(\vec{k'}_\pp)$, which are all
explicitly imposing that energy and the momentum parallel to wall are
conserved quantities.  In the fermionic case we only needed one weight function (a 2x2-density matrix)~\cite{HKR1}. For the present problem we shall need three different functions because of the explicit $k_z$-dependence in eqs.~(\ref{Klein2})-(\ref{Klein3}). Indeed, taking the 0th moment of eq.~(\ref{Klein2}) and the 0th and 1st moments of eq.~(\ref{Klein3}) the following closed set of equations for the moment functions is obtained~\cite{Zhuang}:
\begin{eqnarray}
\frac14 \partial^2_z \rho_0 + k_m^2 \rho_0 - \rho_2 &=& 0
\nonumber\\
\partial_z\rho_1 &=& 0
\nonumber\\
\partial_z\rho_2 - \frac12 (\partial_z k_m^2) \rho_0 &=& 0\,.
\label{rho_Eq1_STA}
\end{eqnarray}
The number of independent moments of course matches the number of independent on-shell functions in the spectral solution. Using eq.~(\ref{fullspec_STA}) we get the following expressions for $\rho_{0,1,2}$ in terms of the on-shell functions $f_{\pm}$ and $f_c$ for $k_0>0$ (and a constant mass $m$):
\begin{eqnarray}
\rho_0 &=& \frac{1}{2k_m}(f_+ + f_-) + f_c 
\nonumber\\
\rho_1 &=& \frac{1}{2}(f_+ - f_-)
\nonumber\\[1mm]
\rho_2 &=& \frac{k_m}{2}(f_+ + f_-)\,,
\label{rho-f_STA}
\end{eqnarray}
while the higher moments are trivially related to $\rho_{1,2}$ by ($n \geq 1$): $\rho_{2n+1} = k_m^{2n}\rho_1$  and $\rho_{2n+2} = k_m^{2n+1}\rho_2$. 
Equations (\ref{rho_Eq1_STA}) are our master equations for solving the Klein problem. It should be noted that we made no approximations to derive them, since all $k$-gradients vanish from eqs.~(\ref{rho_Eq1_STA}) upon integration over $k_z$. Connection formulae (\ref{rho-f_STA}), and the above expressions for the higher moments are formally valid only in the mean field limit. However, they can be used to set the physical boundary conditions between the on-shell functions $f_{\pm,c}$ and moments $\rho_{0,1,2}$ asymptotically in the limit $z \rightarrow \pm \infty$.

It is easy to solve eqs.~(\ref{rho_Eq1_STA}) in the separate regions I
and II for the Klein problem (see figure~\ref{fig:KleinWall}), where the
potential term is either zero or a constant. These solutions will
contain eight unknown constants that can be fixed by the boundary
conditions at $z \rightarrow \pm \infty$ and the matching conditions at the
potential wall $z=0$ induced by the moment equations (\ref{rho_Eq1_STA}). 
First note that all spatial gradients vanish everywhere except at the wall 
at $z=0$. Two latter eqs.~(\ref{rho_Eq1_STA}) then imply that $\partial_z\rho_{1,2} = 0$, \ie $\rho_{1,2}$ are constants in regions I and II. From relations (\ref{rho-f_STA})
it then follows that $f_\pm$ are also constants at $|z| \gg 0$. Now consider the boundary conditions appropriate for our
reflection problem.
The fact that there is no incoming flux from the
left sets $f^{\rm II}_+=0$, because $f_\pm$ were found to be
constants. This condition also sets coherence
solution to zero {\em asymptotically}, \ie $f_c\rightarrow 0$ as
$z\rightarrow-\infty$, since there are no asymptotic mixing states;
note however that we cannot exclude 
coherence at finite distances from the wall based on the boundary
conditions alone. Finally, we can normalize the incoming flux from the
right to unity  $f^{\rm I}_-=1$. After these definitions we have to
account for two distinct possibilities depending on whether the
momentum in the region II, $k_m^{\rm II}$, is real or imaginary.
\FIGURE[t]{\includegraphics[width=12cm]{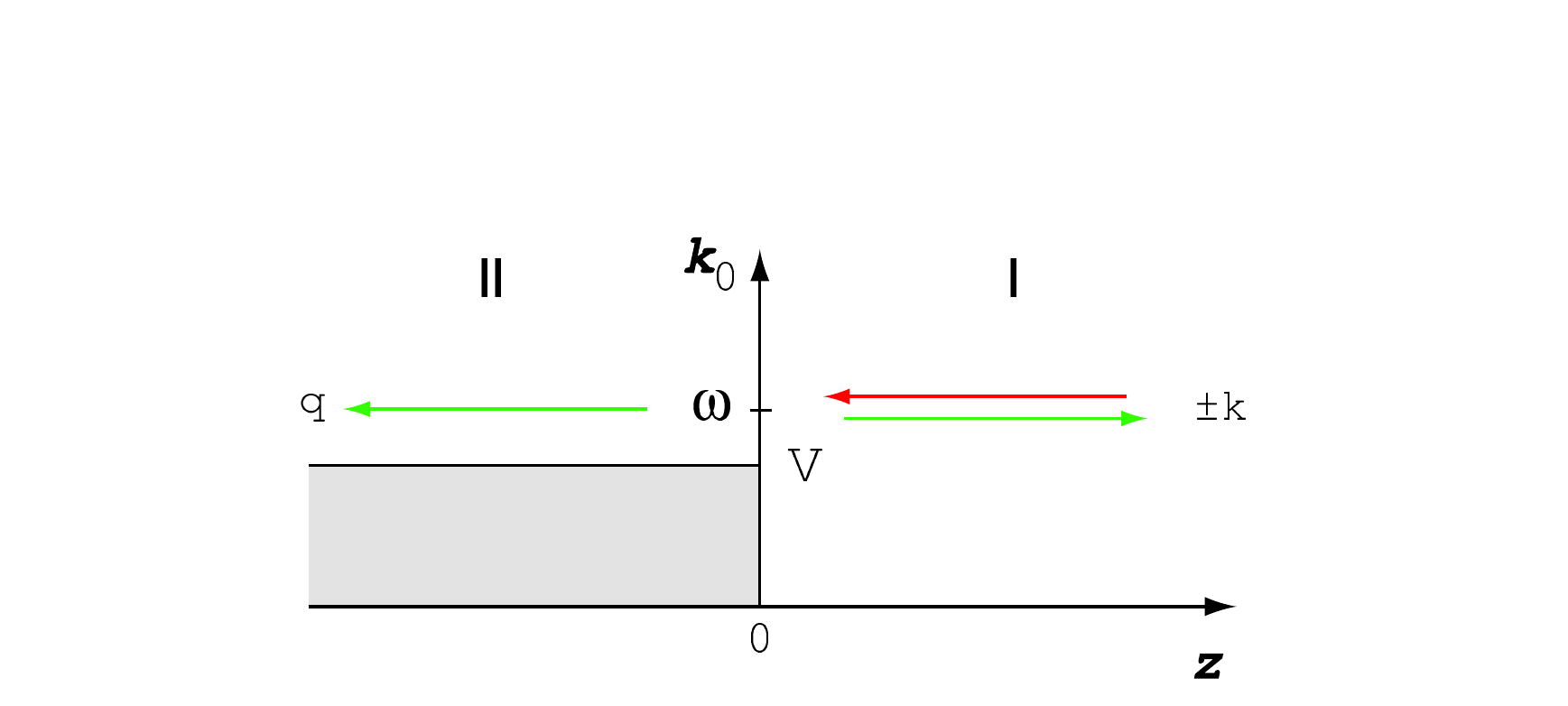}
        \caption{Reflection from a step-like potential. Arrows describe the directions of momenta of the in- and outgoing particles.}
        \label{fig:KleinWall}}
\vskip 0.4truecm

\noindent
Let us first assume that $k_0>V + m$, so that $k_m^{\rm II}\equiv q$ is real.
In this case we can have nonzero transmitted flux $f^{\rm II}_-\neq
0$. Moreover, from the first eq.~(\ref{rho_Eq1_STA}) we find that
$\rho_0$ is oscillatory in both regions I and II ($k_m^{\rm I}\equiv
k$ is always real)
\begin{equation}
\rho^{\rm i}_0 = A^{\rm i} \cos(2k_m^{\rm i}z + \delta^{\rm i}) +
\frac{1}{(k_m^{\rm i})^2} \rho^{\rm i}_2\,,
\label{rho_0_Re}
\end{equation}
where $A^{\rm i}$ and $\delta^{\rm i}$ are new integration constants and
$\rho^{\rm i}_2$ are constants related to $f^{\rm i}_\pm$ by
eq.~(\ref{rho-f_STA}), and we have denoted $i=$I,II. Combining
eqs.~(\ref{rho-f_STA}) with eqs.~(\ref{rho_Eq1_STA}) we find that $f_c
= -\frac{1}{4 k_m^2}\partial_z^2\rho_0$, so that coherence solutions
are also oscillatory. Since coherence should vanish when $z\rightarrow
-\infty$ we find that $A^{\rm II},B^{\rm II} = 0$ in this case. 

The remaining integration constants are fixed by the matching
conditions at $z=0$ induced by the moment equations
(\ref{rho_Eq1_STA}). As $\partial_z\rho_1$ vanishes everywhere we see
that $\rho_1$ must be continuous over the barrier. Equating
$\rho_1^{\rm I} =\rho_1^{\rm II}$ gives the flux conservation
equation:
\begin{equation}
f^{\rm II}_- = 1 - f^{\rm I}_+ \,.
\end{equation}
The last eq.~(\ref{rho_Eq1_STA}) implies that $\rho_2$ must
have a finite discontinuity over the barrier, which can be
computed by integrating it over a step from $z=-\epsilon$ to
$z=\epsilon$ to give $\rho^{\rm I}_2-\rho^{\rm II}_2 =
\shalf[k^2-q^2]\rho_0(z=0)$. Using this in the first
eq.~(\ref{rho_Eq1_STA}) we see that also $\partial^2_z \rho_0$ has at most a finite discontinuity over the barrier implying that
$\rho_0$ and its derivative $\partial_z\rho_0$ are continuous at
$z=0$. Using all these conditions we can fix the remaining integration
constants to eventually find the transmitted and reflected fluxes:
\begin{equation}
f^{\rm I}_+ =\frac{(k-q)^2}{(k+q)^2} \,, \qquad
f^{\rm II}_- = \frac{4 k q}{(k+q)^2} \,,
\end{equation}
which are in accordance with the usual Klein-Gordon approach. Finally, the coherence solution in the region I can be written in the form:
\begin{equation}
f^{\rm I}_c = \frac{1}{k}\sqrt{f^{\rm I}_+} \cos(2 k z)\,.
\end{equation}
It should be noted that if the coherence solution were neglected, the only consistent solution for the reflection problem would have been $f_+^{\rm I} = 0$, \ie that of a classical, complete transmission.

\vskip 0.4truecm

\noindent 
Now consider the case $V-m<k_0<V+m$ for which $q$ is imaginary. In this case we cannot have mass-shell solutions in region II, so that both $f^{\rm II}_\pm=0$. However, we cannot exclude a coherence solution $f_c=-\frac{1}{4 k_m^2}\partial_z^2\rho_0$ as long as it becomes asymptotically zero as $z\rightarrow-\infty$. This indeed turns out to be the case: from eq.~(\ref{rho_Eq1_STA}) we find that $\rho^{\rm II}_{1,2}=0$ and
\begin{equation}
\rho^{\rm II}_0 = A^{\rm II} e^{2|q|z} \,.
\label{rhoII_0_Im}
\end{equation}
In region I the solutions are of the same form as above above with a real $q$. We perform the same matching procedure over the barrier as in the case of real $q$ to fix the values of remaining integration constants. Going through the algebra finally gives the expected result with a complete reflection: $f^{\rm I}_+=1$. The coherence function in the region I is again oscillatory:
\begin{equation}
f^{\rm I}_c = \frac{1}{k}\cos(2 k z + \delta) \quad\quad{\rm with}
\quad \delta = \arcsin\left(\frac{2k|q|}{k^2 + |q|^2}\right)\,,
\end{equation}
while in the region II it is a dying exponential
\begin{equation}
f^{\rm II}_c = \frac{2 k}{k^2 + |q|^2} e^{2|q|z}\,.
\end{equation}
This vanishes as $z\rightarrow -\infty$ as required, but remains nonzero in the vicinity of the wall, where it clearly describes the quantum tunnelling. Since $f_\pm^{\rm II} = 0$, the moment function $\rho_0$ is completely saturated by the coherence function. So the tunnelling effect is a pure coherence phenomenon that can be interpreted as a maximally coherent virtual pair consisting of a left-moving state and its right-moving ``antistate".

\subsection{Particle number}
\label{sec:particle}

As another example, we will consider particle number in a spatially homogeneous system. By taking the real and imaginary parts of equation (\ref{KG_Eq1}) subjected to this particular symmetry we get now:
\begin{eqnarray}
 \Big(k^2 - \frac14 \partial_t^2 - m^2 \cos(\sfrac12
      \partial_t^m \partial_{k_0}^\Delta)\Big)i\prop^<
      &=& 0
\label{KG_Eq_HOM4} \\
 \Big(k_0 \partial_t + m^2 \sin(\sfrac12
      \partial_t^m \partial_{k_0}^\Delta)\Big)i\prop^< &=& 0\,.
\label{KG_Eq_HOM5}
\end{eqnarray}
Analogously to eq.~(\ref{n-moment_Klein}) we again define the $n$-th moments of the Wightman function:
\begin{equation}
\rho_n(|\vec{k}|,t) = \int \frac{{\rm d} k_0}{2\pi}\;k_0^n\,
i\prop^<(k_0,|\vec{k}|,t)\,.
\label{n-moment}
\end{equation}
Again, three lowest moments form a closed set of equations:
\begin{eqnarray}
\frac14 \partial_t^2\rho_0 + \omega_{\vec k}^2 \rho_0 - \rho_2 &=& 0
\nonumber\\
\partial_t\rho_1 &=& 0
\nonumber\\
\partial_t\rho_2 - \frac12 \partial_t(m^2) \rho_0 &=& 0\,.
\label{rho_Eq1_HOM}
\end{eqnarray}
Note that these equations are {\em exact} to all orders of gradients of the
mass $m(t)$, assuming that the surface terms in $k_0$ vanish. Using the full spectral solution eq.~(\ref{fullspec_HOM}) one finds the following relations of moments to  $f_{\pm,c}$:
\begin{eqnarray}
\rho_0 &=& \frac{1}{2\omega_{\vec k}}(f_+ - f_-) + f_c 
\nonumber\\
\rho_1 &=& \frac{1}{2}(f_+ + f_-)
\nonumber\\[1mm]
\rho_2 &=& \frac{\omega_{\vec k}}{2}(f_+ - f_-)\,.
\label{rho-f_HOM}
\end{eqnarray}
Unlike the evolution equations, these relations are only valid in the mean field limit. Moreover, in this approximation, the higher moments are again related to $\rho_{1,2}$ by ($n \geq 1$): $\rho_{2n+1} = \omega_{\vec k}^{2n}\rho_1$ and $\rho_{2n+2} = \omega_{\vec k}^{2n+1}\rho_2$. Relations (\ref{rho-f_HOM}) can be inverted to give $f_{\pm,c}$ in terms of $\rho_i$, $i=0,1,2$. Following the
Feynman-St\"uckelberg interpretation we now {\em define} the phase space number densities for particles and for antiparticles respectively as 
\begin{equation}
  n_{\vec{k}} \equiv f_+(|\vec{k}|) \quad {\rm and} \quad 
  \bar{n}_{\vec{k}} \equiv -1-f_-(|\vec{k}|) \,.
\end{equation}
In terms of the three independent momentum components $\rho_i$, $i=0,1,2$ we then find:
\begin{eqnarray}
n_{\vec{k}} &=& \frac{1}{\omega_{\vec k}}\rho_2 + \rho_1
\nonumber\\
\bar{n}_{\vec{k}} &=& \frac{1}{\omega_{\vec k}}\rho_2 - \rho_1 - 1
\nonumber\\
f_c(|\vec{k}|) &=& \rho_0 - \frac{1}{\omega_{\vec k}^2}\rho_2 \,.
\label{f-rho_HOM}
\end{eqnarray}
Using the free theory equations of motion (\ref{rho_Eq1_HOM}) we can solve: $\rho_2 = \frac14 \partial_t^2\rho_0 + \omega_{\vec k}^2 \rho_0$ to get the expression for the particle number in the form
\begin{equation}
n_{\vec{k}} = \omega_{\vec k} \rho_0 +
\frac{1}{4\omega_{\vec k}}\partial_t^2\rho_0 + \rho_1 
\label{part_number}
\end{equation}
and the coherence solution now becomes just $f_c(|\vec{k}|) = -\partial_t^2\rho_0/(4\omega_{\vec k}^2)$.  Note that the moment $\rho_1$ remains a constant in a free theory, where $\partial_t\rho_1=0$ by eq.~(\ref{rho_Eq1_HOM}). Setting a constraint $\rho_1 = -1/2$ then fixes $n_{\vec{k}}=\bar{n}_{\vec{k}}$ at all times, consistent with the fact that $\phi$ is a real scalar field. In the operator formalism this constraint is imposed  by the Wronskian normalization of the mode functions~\cite{Prokopec_partnumber}. These results extend trivially to the case of a complex scalar field; the only difference is that then $\bar n_{\vec k}$ can differ from $n_{\vec k}$ and $\rho_1$ becomes a free parameter related to the chemical potential.

Let us now compare our particle number (\ref{part_number}) with other
definitions in the literature. Taking into account the free theory
equation of motion (\ref{rho_Eq1_HOM}) (and the thermal form of the spectral function, eq.~(\ref{spectral_function}) below), the definition of the particle number by Aarts and Berges in ref.~\cite{Berges} can be expressed as
\begin{equation}
\Big(n_{\vec{k}} + \frac12\Big)^2 = \rho_0\Big(\omega_{\vec k}^2\rho_0 +
\frac12 \partial_t^2 \rho_0 \Big)\,.
\label{Berges}
\end{equation}
This agrees with our result (\ref{part_number}) in the adiabatic, or small coherence limit: $\partial_t^2 \rho_0 << \omega_{\vec k}^2\rho_0$, as can be directly seen by solving $n_{\vec{k}}$ and expanding the square root in eq.~(\ref{Berges}).  

One interesting application is to consider particle number evolution
during inflation. When applied to expanding space-times in conformal
coordinates all that changes in previous equations is replacing time
with a conformal time: $t \rightarrow \eta$ and the mass by an
effective mass $m^2 \rightarrow \bar{m}^2 = a^2 m^2 - \partial_\eta^2
a /a$, where $a(\eta)= -1/(H\eta)$, $\eta<0$, is the scale factor. Matching with the Bunch-Davies vacuum at early times, one finds that during pure De Sitter phase
\begin{equation}
\rho_0 = - \frac{1}{4}\pi\eta |H^{(1)}_\nu(-k\eta )|^2 
\quad \stackrel{m=0}{\rightarrow} \quad \frac{1}{2k}\Big(1+\frac{1}{(k\eta)^2}\Big) 
\,,
\label{rho0infl}
\end{equation}
where $H_\nu^{(1)}$ is the Hankel function of the first kind with $\nu^2 \equiv 9/4 -(m/H)^2$ and $H$ is the Hubble expansion rate. It is easy to see that the gradient expansion in the De Sitter case can be rewritten as an expansion in $1/|k\eta |$, so that our inversion formulae (\ref{rho-f_HOM}) provide a good approximation at early times, but break at the horizon crossing at $k\eta \approx 1$.  Using eq.~(\ref{part_number}), still with $\rho_1 \equiv -1/2$ we find that at early times, or ultraviolet limit $k/a \gg H$, our particle number behaves as
\begin{equation}
n_{\vec{k}} \approx 16 a^6 \Big(Ê\frac{H}{2k}\Big)^6 \,.
\label{inflationpartnum}
\end{equation}
This result differs from the particle number defined in ref.~\cite{Prokopec_partnumber}:
\begin{equation}
n_{\vec{k}} \equiv \omega_{\vec k} \rho_0 +
\frac{1}{4\omega_{\vec k}} \partial_\eta^2\rho_0 - \frac12 -
\frac{1}{2\omega_{\vec k}} \frac{d}{d\eta}\left(\frac{\partial_\eta a}{a}
    \rho_0 \right)\,,
\label{Prokopec}
\end{equation}
which at early times times becomes $n_{\vec{k}} \approx  a^2 (H/2k)^2$. This is not really surprising, because the particle number is not unambiguously defined in curved spacetimes. The particle number (\ref{Prokopec}) is found by a  diagonalization of the Hamiltonian and it corresponds to a maximum particle number seen by an ideal detector. Our definition relies on phase space arguments and corresponds to an {\em adiabatic} particle number~\cite{BirrelAndDavis}, which rather tries to minimize $n$. In particular for a conformally coupled scalar theory our particle number can be shown to remain zero at all times if it was set to zero in the beginning.

\subsection{Energy density and pressure}

Let us next compute the expectation values of energy (Hamiltonian)
density and pressure, which are the $00$- and the $ii$-components of
the energy-momentum tensor ($g^{\mu\nu}$ is the standard Minkowskian
metric tensor with the signature $(+,-,-,-)$):
\begin{equation}
T^{\mu\nu}=\frac{\partial{\cal
  L}}{\partial(\partial_\mu\phi)}\partial^\nu\phi - {\cal
  L}g^{\mu\nu}\,,
\end{equation}
respectively, in the spatially homogeneous case. For the energy
density we get using the free theory equations of motion
(\ref{KG_Eq_HOM4})-(\ref{KG_Eq_HOM5}) 
\begin{eqnarray}
\langle{\cal H}(t)\rangle = \langle T^{00}(t)\rangle 
&=&  \langle \sfrac12(\partial_t \phi)^2+ \sfrac12(\partial_{\bf x} \phi)^2 + \sfrac12 m^2Ê\phi^2 \rangle
\nonumber\\
&=& 
\int \frac{{\rm d}^4 k}{(2\pi)^4}\,k_0^2
i\prop^<(k_0,|\vec{k}|,t) 
\nonumber\\
&=& \int \frac{{\rm d}^3 k}{(2\pi)^3}\,\frac12(n_{\vec{k}} + \bar{n}_{\vec{k}}+1) \,\omega_{\vec k} \,, 
\label{edensity}
\end{eqnarray}
and for the pressure we get in the same way
\begin{eqnarray}
\langle P(t)\rangle = \langle T^{ii}(t)\rangle 
&=& \int \frac{{\rm d}^4 k}{(2\pi)^4}\,\Big(\frac13 \vec{k}^2 + k_0^2-\omega_{\vec k}^2 \Big)
i\prop^<(k_0,|\vec{k}|,t) 
\nonumber\\
&=& \int \frac{{\rm d}^3 k}{(2\pi)^3}\,\left[\frac{\vec{k}^2}{3 \omega_{\vec k}}
\frac12(n_{\vec{k}} + \bar{n}_{\vec{k}}+1) - \Big(\omega_{\vec k} - \frac{\vec{k}^2}{3\omega_{\vec k}}\Big) \omega_{\vec k}f_c \right]\,. 
\label{pdensity}
\end{eqnarray}
There is an explicit contribution from the coherence shell function $f_c$ in the pressure, signalling that at the quantum level pressure differs from
the statistical one. However, as discussed in the analysis with fermions \cite{HKR2}, we expect that in most cases the direct coherence contribution is unobservable in the time-scales longer than the typical oscillatory time $\Delta t_{\rm osc} \sim 1/\omega$ of the coherence solution. However if a strong amplification mechanism is in place as in during inflation, even these coherent small scale oscillations might have physical consequences.  

Let us point out one delicate issue in computing the energy density and pressure from the 2-point function. For example our result for the energy density (\ref{edensity}) follows directly from the direct space integral expression for the Hamiltonian. However, one might instead try to start from a partially integrated form of the Hamiltonian density:
\begin{equation}
\tilde{\cal H} \equiv \sfrac12 
\phi \left( -\partial_t^2 - \partial_{\bf x}^2 + m^2Ê\right) \phi 
\label{newhami}
\end{equation}
Normally this Hamiltonian would give same result as ${\cal H}$, since the only effect of using the form (\ref{newhami}) in (\ref{edensity}) would be replacing $k_0^2 $ by $\omega_{\vec k}^2$ in the expression in the second line. Here this difference matters however, because in the former case the $k_0$-shell does not contribute to the energy density whereas in the latter it does. This observation  nicely underlines the nonlocal nature of our coherence solutions. Indeed, although our equations are parametrized by a well defined external time variable, the shell $k_0= 0$ corresponds to a completely delocalized, constant mode in the {\em internal} variable $u-v$ of the 2-point function $\langle \phi(v) \phi(u)\rangle$.  Thus, the partial integration and the subsequent neglecting of the boundary term leading to the alternative Hamiltonian $\tilde {\cal H}$  is not a legitimate operation here.

\section{Non-relativistic case, Schr\"odinger equation}
\label{sec:non-relativistic}

Our methods can also be applied to non-relativistic problems. The extension is very simple, and we give it here for completeness for a field $\psi$, possibly interacting with some background potential $V$. That is, assume that $\psi$ obeys the Schr\"odinger equation 
\begin{equation}
i\partial_t \psi = H \psi = \left(-\frac{1}{2m}\nabla^2 + V\right)\psi\,. 
\end{equation}
The Wightman function $i\prop_\psi^<(u,v) \equiv \langle  \psi^\dagger(v)\psi(u)\rangle$ then obeys the equation:
\begin{equation}
\Big(i\partial_{u^0} + \frac{1}{2m}\nabla_u^2 - V(u)\Big)i\prop_\psi^<(u,v) = 0\,,  
\end{equation}
which in the mixed representation reads:
\begin{equation}
\Big(k_0 - \frac{\vec{k}^2}{2m} + \frac{1}{8m}\nabla_x^2 +
  \frac{i}{2}\partial_t + \frac{i}{2m}\vec{k}\cdot\vec{\nabla} -
  V(x) e^{-\frac{i}{2} \partial^V_x \cdot \partial^\prop_k} \Big)i\prop^<_\psi(k,x) = 0\,.   
\label{non-rel1}
\end{equation}
This equation resembles the free dynamical equation (\ref{KG_Eq1}) for a relativistic scalar field. The physical content of eq.~(\ref{non-rel1}) can again be most easily analyzed by studying the spatially homogeneous and the static, planar symmetric cases. 

In the homogeneous case we assume that the potential and the solutions $\prop^<_\psi$ can only depend on time, so that:
\begin{eqnarray}
 \Big(k_0 - \frac{\vec{k}^2}{2m} - 
  V(t)\cos(\sfrac12 \partial^V_t \partial^\prop_{k_0})\Big)i\prop^<_\psi(k,t) &=& 0
\label{KG_Eq_HOM2_b}\\
 \Big(\frac12\partial_t + V(t)\sin(\sfrac12 \partial^V_t \partial^\prop_{k_0})\Big)i\prop^<_\psi(k,t) &=& 0\,.
\label{KG_Eq_HOM3_b}
\end{eqnarray}
This set of equations clearly has only positive energy particle solutions with the non-relativistic dispersion relation $k_0 = \frac{\vec{k}^2}{2m} + V$ in the mean field limit, but no negative energy antiparticle solutions, nor any coherence solution living at $k_0=0$. This was to be expected because antiparticles are not automatically a part of the spectrum of a non-relativistic field theory. The absence of $k_0=0$-shell here is thus consistent with its interpretation as describing particle-antiparticle coherence. In static, planar symmetric case one still finds the $k_z=0$-shell solution, which describes the spatial reflection coherence in accordance to relativistic fields.

\subsection{Planar symmetric problems and bound states}

Static, planar symmetric case is more interesting in the
nonrelativistic limit. Here the equations become identical to
eqs.~(\ref{Klein2})-(\ref{Klein3}) for the relativistic scalar field in
section \ref{sec:planar}, apart from mass-shell dispersion relation,
which here reads:
\begin{equation}
k_m=\pm\sqrt{2m(k_0 - V) - \vec{k}_\pp^2} \,.
\end{equation}
This is of course just the nonrelativistic limit of the dispersion relation used in (\ref{Klein2})-(\ref{Klein3}). Moreover, the moment equations
are also obviously identical in form to equations
(\ref{rho_Eq1_STA}). 

As an example, let us consider the familiar infinite square well potential in
1-dimensional quantum mechanics. Solving the moment equations 
(\ref{rho_Eq1_STA}) and imposing the same matching conditions on the
well boundaries as for the Klein problem, one finds that the solution
consistent with the asymptotic vanishing of $\rho_0$ is
\begin{eqnarray}
\rho_0 &=& \theta_{\rm II}(z) \sum_{n=1}^\infty A_n \frac{2\pi}{L}\Big(1 +
  (-1)^{n+1}\cos(2 k_n z)\Big) \delta(k_0-E_n) 
\nonumber \\
\rho_1 &=& 0 
\nonumber \\
\rho_2 &=& \theta_{\rm II}(z) \sum_{n=1}^\infty A_n \frac{2\pi}{L}
k_n^2\,\delta(k_0-E_n) \,,
\label{nonrel-rho}
\end{eqnarray}
where the mass-shell momentum and energy are quantized: $k_m = k_n
\equiv \pi n/L$ and $E_n \equiv k_n^2/(2m)$, with $n=1,2,\ldots$. The function
$\theta_{\rm II}(z) \equiv \theta(z+L/2) - \theta(z-L/2)$ restricts
the solution to be nonzero only inside the potential well. The
remaining constants $A_n$ can be set by the normalization of the
solution. We shall soon see that for a pure state these constants will
correspond to the occupation numbers of the one-particle states
labelled by quantum number $n$. Next we shall interpret these results in terms of the spectral shell solutions. The generic (mean field) spectral solution for the problem is given by
\begin{equation}
i\prop^<_\psi = 2\pi \left[ 2m f_{s_{k_z}} \delta(k_z^2 - k_m^2) + f_c \delta (k_z) \right] \,.
\end{equation}
With this normalization the relations between $f_i$ and the moments read:
\begin{eqnarray}
\rho_0 &=& \frac{m}{k_m}(f_+ + f_-) + f_c 
\nonumber\\
\rho_1 &=& m \,(f_+ - f_-)
\nonumber\\[1mm]
\rho_2 &=& mk_m \, (f_+ + f_-)\,.
\label{rho-nonrel_STA}
\end{eqnarray}
From equations (\ref{nonrel-rho})-(\ref{rho-nonrel_STA}) one finds now 
\beq
i\prop^<_\psi = \theta_{\rm II}(z) \sum_{n=1}^\infty A_n
\frac{2\pi}{L}\Big[ k_n \delta(k_z^2 - k_n^2) + (-1)^{n+1}\cos(2 k_n
  z)\delta (k_z)\Big] 2\pi \delta(k_0 - E_n)\,.
\label{spec_correlator}
\eeq
We can estimate how good approximation this singular shell picture is
by computing the correlator $i\prop^<_\psi$ directly from the
one-particle wave functions. For that, we consider a pure state
\begin{equation}
| \xi \rangle \equiv | 1^{f_1}, 2^{f_2}, ... \rangle \equiv \prod_i \frac{(\hat a_i^\dagger)^{f_i}}{\sqrt{f_i!}} |0 \rangle \,,
\end{equation}
where $[\hat a_n,\hat a_m^\dagger ]=\delta_{n,m}$. The Wightman
function corresponding to this state is
\begin{equation}
i\prop^<_{\psi,\xi} \equiv \langle \xi |\hat \psi^\dagger (t,z)\hat \psi(t',z')| \xi\rangle \,,
\label{state_correlator}
\end{equation}
\FIGURE[t]{\includegraphics[width=12.5cm]{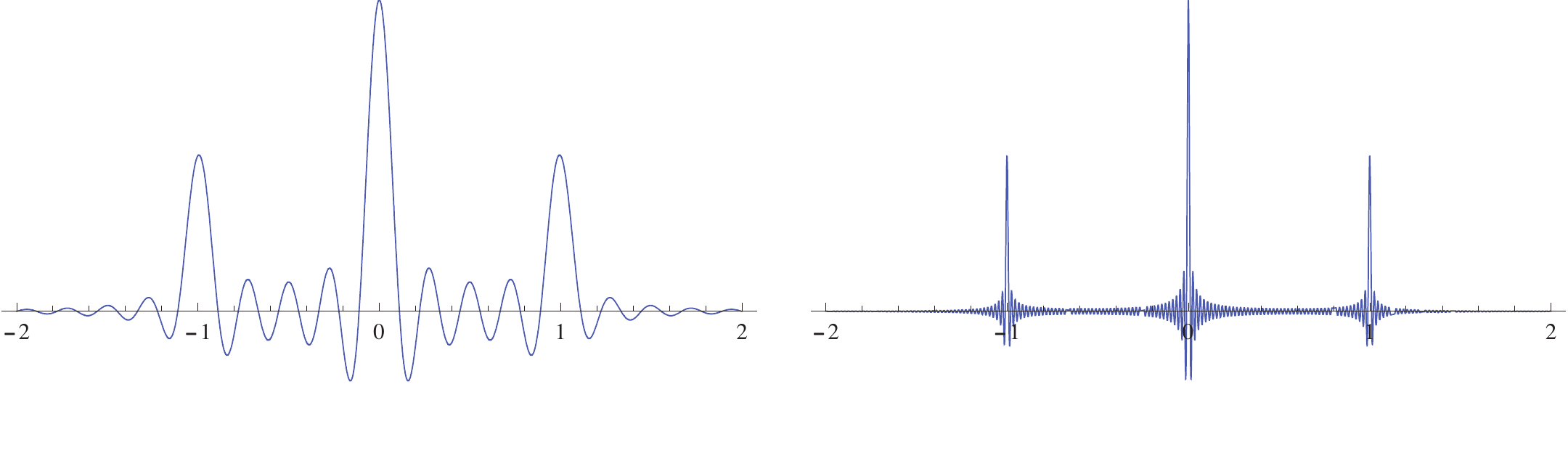}
        \caption{The phase space structure of the correlator
       $i\prop^<_{\psi,\xi}(k,z)$. Shown is the expression
       inside the square brackets in eq.~(\ref{mom_spread_correlator})
       as a function of $k_z/k_n$ at $z=0$ for $n=9$ (left) and $n=99$
       (right).}
        \label{fig:shells}}
\noindent
where $\hat \psi(t,z) = \sum_n e^{-iE_nt}\psi_n(z)\hat a_n$ is the field
operator and 
\beq
\psi_n(z) = \left\{ \begin{array}{ll} 
    \sqrt{2/L}\cos(k_n z) \theta_{\rm II}(z)\,, & \quad n=1,3,5,\ldots \\[1mm]
    \sqrt{2/L}\sin(k_n z) \theta_{\rm II}(z)\,, & \quad n=2,4,6,\ldots
    \end{array} \right.
\eeq
are the normalized one-particle wave-functions in the well. A direct computation of the zeroth moment then gives exactly the same expression as in eq.~(\ref{rho-nonrel_STA}) with $A_n = f_n$. However,  the full Wigner transformed correlator (\ref{state_correlator}) becomes now:
\beqa
     &&i\prop^<_{\psi,\xi}(k,z) = \theta_{\rm II}(z) \sum_{n=1}^\infty f_n
     \frac{2\pi}{L}\bigg[ \sum_\pm \frac{\sin((k_z \mp k_n)(L-2|z|))}{k_z \mp
      k_n} + \nonumber\\
     &&\qquad\qquad\qquad\qquad\qquad\quad\,
     (-1)^{n+1}\frac{\sin(k_z(L-2|z|))}{k_z}2\cos(2 k_n z)\bigg]\delta(k_0 - E_n)\,.
\label{mom_spread_correlator}
\eeqa
We can see that the entire $k_z$-dependence of this correlator is encoded
into a function $\sin(q(L-2|z|))/q$, which is a representation of the
Dirac delta function in the limit $k_n(L-2|z|) \gg 1$:
\beq
\sin(q(L-2|z|))/q \longrightarrow \pi \delta(q)\,.
\eeq
In this limit the correlator (\ref{mom_spread_correlator}) reduces to
the spectral form shown in eq.~(\ref{spec_correlator}). In
figure~\ref{fig:shells} we have plotted the correlator (the expression
in the square brackets) eq.~(\ref{mom_spread_correlator}) as a
function of $k_z$ at the centre of the square well at $z=0$ for $n=9$
and $n=99$ corresponding to $k_n(L-2|z|) \approx 30$ and $300$,
respectively. It is clear that the phase space structure approaches
the singular form as the momentum scale $k_n$, compared to the
distance from the walls, increases. Similar conclusions hold also in
the case of mass- or potential walls. In particular for a step wall
the spectral form for the correlator becomes exact when the distance
of the wall is large in units $1/k$.

Let us stress that our only use for the singular shell approximation
is to relate the moment functions to on-shell functions in adiabatic regime 
and to provide a practical scheme to evaluate the collision term that
gives rise to a closed set of equations for the moments $\rho_{0,1,2}$
(or equivalently the on-shell functions $f_{\pm,c}$). The results of
this section show that that this scheme can be useful even in most
extreme situations. Indeed, even a correlator shown in the left panel
of figure~\ref{fig:shells} should be reasonably well represented by a
singular ansatz in the collision integral whenever the matrix element
is a relatively smooth function of momentum. For smooth wall profiles
and for slowly varying driving forces the correlator can approach the
singular ansatz even inside the transition region. The quantitative
measure for this is the ratio of the momentum to the rate of change of
the potential. Finally, let us point out that our results show that
the new $k_z=0$-shell is equally well visible and concentrated as the
standard mass-shells. That is, our approximation for the phase
space is in no means less rigorous than the standard derivation of
the ordinary Boltzmann equations relying only on the mass-shell
contributions.

\section{Spectral function and thermal limit}
\label{sec:spectral}

Having shown that the coherence solutions are part of the spectrum of the dynamical 2-point function, let us now show that they do {\em not} appear in the pole functions, or in particular in the spectral function ${\cal A}$. We shall only consider the spatially homogeneous noninteracting case. With no interactions the  equation for ${\cal A}$ eq.~(\ref{SpecEqMix1}) becomes identical with the K-G equation (\ref{KG_Eq1}) for $\prop^<$. Consequently, the most general solution fulfilling the mean field phase space constraints must be identical to eq.~(\ref{fullspec_HOM}) with three yet undefined on-shell functions $f^{\cal A}_{\pm}$ and $f^{\cal A}_c$. However, the spectral function must also satisfy the {\em spectral sum-rule}, which in direct space reads
\begin{equation}
2\,\partial_{t'}{\cal A}(t',\vec{u}; t,\vec{v})_{t'=t} =
-i\delta^3(\vec{u}-\vec{v})\,.
\label{sumrule1}
\end{equation}
This follows from the canonical equal time commutation relations of the scalar field $\phi$, or it can be derived from the pole equations (\ref{SpecEqMix1})-(\ref{SpecEqMix2}). Transforming
eq.~(\ref{sumrule1}) to the mixed representation gives
\begin{equation}
\int \frac{{\rm d}k_0}{\pi}\Big(k_0 + \frac{i}{2}\partial_t\Big){\cal
  A}(k,x) = 1\,.
\label{sumrule2}
\end{equation}
The time-derivative appearing in this representation is usually omitted in the literature, by an implicit assumption of translational invariance. This is appropriate for example for thermal equilibrium systems, but for more general nonequilibrium problems it does give a new independent constraint. In terms of moment functions $\rho^{\cal A}_n\equiv \int \frac{{\rm d} k_0}{2\pi}\;k_0^n\, {\cal A}$ the sum rule (\ref{sumrule2}) becomes:
\begin{equation}
\rho^{\cal A}_1 = \frac12 \,,\qquad \partial_t\rho^{\cal A}_0 = 0\,.
\end{equation}
The latter constraint implies that $\rho^{\cal A}_0 = {\rm const.}$ and  furthermore $\partial_t^2\rho^{\cal A}_0 = 0$. Dynamical equations for the moment functions $\rho^{\cal A}_n$ are of course identical to eqs.~(\ref{rho_Eq1_HOM}) for $\rho_n$. Imposing the sum-rule constraints on these equations we find that also $\rho^{\cal A}_2 = {\rm const.}$  One then finds that either $\rho^{\cal A}_0=\rho^{\cal A}_2 \equiv 0$, or the mass is a constant everywhere. To get a continuous constant mass limit we must always set: 
\begin{equation}
\rho^{\cal A}_0=\rho^{\cal A}_2=0\,,\qquad \rho^{\cal A}_1 = \frac12\,.
\end{equation}
Connection formulae identical to eq.~(\ref{rho-f_HOM}) between $\rho^{\cal A}_n$ and $f^{\cal A}_{\pm,c}$ then give:
\begin{equation}
f^{\cal A}_\pm = \frac12\,, \qquad f^{\cal A}_c = 0\,.
\end{equation}
These will finally reduce ${\cal A}$ to its standard thermal form
\begin{equation}
{\cal A} = \pi{\rm sgn}(k_0)Ê\, \delta \left(k^2 - m^2\right)\,.
\label{spectral_function}
\end{equation}
%

\noindent
Let us finally consider the thermal equilibrium limit for the function $i\prop^<$. The new constraining element here is the Kubo-Martin-Schwinger (KMS) boundary  condition 
\begin{equation}
\prop^>(t) =  \prop^<(t +i\beta ) \,.
\end{equation}
However, first note that the relation $\prop^> - \prop^< = - 2i {\cal A}$ sets
\begin{equation}
f^>_\pm - f^<_\pm = 1 \quad {\rm and} \quad f^>_c - f^<_c = 0 \,.
\label{relationshere}
\end{equation}
Then the momentum space version of the KMS-condition: $\prop^>_{\rm eq}(k_0) = e^{\beta k_0}\prop^<_{\rm eq}(k_0)$ is enough to set the mass-shell distributions to the statistical limit:
\begin{equation}
f^<_{k_0} = n_{\rm eq}(k_0) \quad {\rm and} \quad 
f^>_{k_0} = 1 + n_{\rm eq}(k_0)  \,,
\label{eqfspm}
\end{equation}
where $n_{\rm eq}(k_0) = 1/(e^{\beta k_0} -1)$ is the usual Bose-Einstein distribution. Moreover, coherence functions are subjected to constraint
\begin{equation}
f^<_c = f^>_c \,.
\label{eqf12}
\end{equation}
However, the KMS-condition only makes sense when the system has a time-independent Hamiltonian. This implies time-translational invariance in real time which immediately eliminates coherence, leading to the standard thermal expressions~\cite{PSW}:
\begin{eqnarray}
i\prop^<_{\rm eq} &=& 2\pi{\rm sgn}(k_0) n_{\rm eq}(k_0) \delta\left(k^2
  - m^2\right)\,,
\nonumber \\
i\prop^>_{\rm eq} &=& 2\pi{\rm sgn}(k_0) (1+n_{\rm eq}(k_0)) \delta\left(k^2
  - m^2\right)\,.
\label{thermal_prop}
\end{eqnarray}

\section{The case with collisions}
\label{sec:interactions}

We now move to consider the case with collisions. As explained in the
introduction we are using exact forms of the integrated evolution
equations except for the evaluation of the collision terms. It is only
there that we need to use the (in general approximate) connection
formulae (\ref{rho-f_HOM}) between the moments and the on-shell
functions to get the collision integrals in closed form. The complete
set of Schwinger-Dyson equations of the interacting theory (\ref{SpecEqMix1})-(\ref{DynEqMix}) are too complicated to be used in practical applications without approximations. Here we make a series of approximations that will enable us to consider the essential quantum dynamics in terms of the three lowest moments $\rho_{0,1,2}$ in the presence of collisions. 

First, we will consider a weakly interacting theory, so that the usual {\em quasiparticle} approximation applies. This means that the term $\propto \Pi_H \prop^<$ is included to modify the dispersion relations in both the pole equations (\ref{SpecEqMix1})-(\ref{SpecEqMix2}) and in the dynamical equation (\ref{DynEqMix}) for the Wightman function $i\prop^<$. However the terms $\propto \Gamma$, which are the source of broadening of the phase space of the pole functions are neglected in the pole equations entirely to allow a singular phase space structure, as usual in thermal field theory. Using the constraint $\prop^>-\prop^< = -2i{\cal A}$ one can show that within the quasiparticle approximation it is consistent to drop the term $\propto \Pi^< \prop_H$ in the dynamical equation (\ref{DynEqMix}), and neglect the collision term when working out the spectral structure for $\prop^<$ , even if in the coupling constant expansion the dropped terms are of same order as $ \Pi_H$ and collision integral $\cal C $. A more complete derivation of these approximations and discussion of the role of different self-energy functionals could be found in ref.~\cite{PSW}.
Second, we will compute the collision term in the r.h.s. of eq.~(\ref{DynEqMix}) only up to first order gradients; this should be a good approximation at least for cases where the variations are affecting only a small subset of the entire interacting system. With these approximations one can find the desired reduction to three moments only. However, we shall here neglect also the term $\sim \Pi_H \prop^<$, which would just change the dispersion relations of the states, without altering the qualitative aspects of collisions on the evolution of the system. This leaves us with the flow term of the free theory. The final form of the dynamical equation with collisions in the spatially homogeneous case then reads:
\begin{equation}
 \Big(k^2 - \frac14 \partial_t^2 + ik_0 \partial_t - m^2 e^{-\sfrac{i}{2}
      \partial_t^m \partial_{k_0}^\Delta}\Big)i\prop^<
  = i{\cal C}_{\rm coll},
\label{DynEqMixQPA}
\end{equation}
where the collision term is
\begin{eqnarray}
{\cal C}_{\rm coll} 
&=&  -\Big(\Gamma\,i\prop^< - i\Pi^<{\cal A}\Big) + i\Diamond\Big(
  \{\Gamma\}\{i\prop^<\} - \{i\Pi^<\}\{{\cal A}\}\Big)
\nonumber\\
&\equiv& {\cal C}_0 + i{\cal C}_1\,.
\label{collintegral_first}
\end{eqnarray}
Taking the real and imaginary parts of eq.~(\ref{DynEqMixQPA}) gives the coupled equations:
\begin{eqnarray}
 \Big(k^2 - \frac14 \partial_t^2 - m^2 \cos(\sfrac{1}{2}
      \partial_t^m \partial_{k_0}^\Delta)\Big)i\prop^<
      &=& -{\cal C}_1
\nonumber\\
 \Big(k_0 \partial_t + m^2 \sin(\sfrac{1}{2}
      \partial_t^m \partial_{k_0}^\Delta)\Big)i\prop^<
      &=& {\cal C}_0\,.
\label{KG_Eq_Coll1}
\end{eqnarray}
Our first task is to find the singular shell structure for $i\prop^<$. As
explained above, consistency with the quasiparticle limit of the phase
space requires neglecting the collision terms and also the gradients
of the mass $m^2$ (as explained in section \ref{sec:shell}), so that we are left with the same mean field
constraint equations (\ref{KG_Eq_HOM2})-(\ref{KG_Eq_HOM3}) as in the
free theory case. We then find the familiar shell structure for the Wightman function:
\begin{equation}
i\prop^< = 2\pi\left({\rm
    sgn}(k_0)f_{s_{k_0}}\delta(k^2 -m^2) + f_c\delta(k_0)\right) \,,
\label{sinkku}
\end{equation}
and the same relations between the functions $f_{\pm,c}$ and the moments $\rho_{0,1,2}$ as in the free field case, given by eq.~(\ref{rho-f_HOM}). These relations are the core of our approximation scheme, since they allow the equations of motion derived from (\ref{KG_Eq_Coll1}) to  close with only the three lowest moments.  Indeed, integrating both equations in (\ref{KG_Eq_Coll1}) with a flat weight and the second equation weighted by $k_0$, we get the following generalizations of the free-field moment equations (\ref{rho_Eq1_HOM}):
\begin{eqnarray}
\frac14 \partial_t^2 \rho_0 + \omega_{\vec k}^2 \rho_0 - \rho_2 &=& -\left<{\cal C}_1\right>
\nonumber\\
\partial_t\rho_1 &=& \left<{\cal C}_0\right>
\nonumber\\
\partial_t\rho_2 - \frac12 \partial_t(m^2) \rho_0 &=&  \left<k_0{\cal C}_0\right>\,,
\label{rho_Eq_Coll1}
\end{eqnarray}
where the collision integrals appearing on the r.h.s. of eqs. (\ref{rho_Eq_Coll1}) are
\begin{eqnarray}
\left<{\cal C}_1\right> 
&=& \frac12\partial_t \, \int \frac{{\rm d}
    k_0}{2\pi}\left(\partial_{k_0}\Gamma\,i\prop^<
    - \partial_{k_0}i\Pi^<\,{\cal A}\right)
\nonumber\\
 \left<{\cal C}_0\right> 
&=& -\int \frac{{\rm d} k_0}{2\pi} \left(\Gamma\,i\prop^< - i\Pi^<{\cal A}\right)
\nonumber\\
\left<k_0{\cal C}_0\right> 
&=& -\int \frac{{\rm d} k_0}{2\pi} k_0 \left(\Gamma\,i\prop^< - i\Pi^<{\cal A}\right) \,.
\label{coll_terms_thermal1}
\end{eqnarray}
The problem with these equations is that functions $\Gamma$ and $\Pi^<$ can have an arbitrary phase space structure, so that the collision integrals are a priori not related to the moments $\rho_i$ in any simple way. That is, equations (\ref{rho_Eq_Coll1})-(\ref{coll_terms_thermal1}) do not close. This is of course to be expected, because integration erases a lot of information from the system. Equations (\ref{rho_Eq_Coll1})-(\ref{coll_terms_thermal1}) are in fact useful  only if the collision terms can be reasonably well approximated by some expansion in the lowest moments. This is precisely what our singular shell structure for the Wightman function $i\prop^<$ does. Indeed, when the structure (\ref{sinkku}) is fed into the collision integrals (\ref{coll_terms_thermal1}), they become completely parametrized by the on-shell functions $f_{\pm,c}$, which on the other hand are related to the lowest moments $\rho_{0,1,2}$ via eq.~(\ref{rho-f_HOM}). Note that this approach is more elaborate than a simple truncation of the moment expansion, because the singular shell structure provides nontrivial information about the phase space of the collision integrals. This is of particular importance for the coherence shells, as we shall see below.

To be specific, let us assume a simple thermal interaction for which
the self-energies do not depend on $\prop^<$ and obey the KMS-relation
$\Pi^>=e^{\beta k_0}\Pi^<$. Moreover, it is natural to require (at
least in the vicinity of the mass-shell) that
$\Gamma(-k_0)=-\Gamma(k_0)$ and  $\Gamma(k_0=0)=0$.  These assumptions
should hold quite generally for a thermal $\Gamma$; in the appendix
\ref{sec:self-energy} we will show explicitly that they hold in the
case of a three body Yukawa interaction. Then, using the relation $i\prop^<_{\rm eq} = 2 n_{\rm eq} {\cal A}$ given by eqs.~(\ref{spectral_function}) and (\ref{thermal_prop}), and the inverse relations of eq.~(\ref{rho-f_HOM}) we find:
\begin{eqnarray}
\left<{\cal C}_1\right> 
&=& \frac12 \partial_t
\left[\frac{1}{2\omega_{\vec k}}\partial_{k_0}\Gamma_m(f_+ - f_-)
  +\partial_{k_0}\Gamma_0 f_c\right] 
\nonumber\\
 \left<{\cal C}_0\right> 
&=& - \frac{1}{2\omega_{\vec k}}\Gamma_m\left[(f_+ + f_-) - (f^{\rm eq}_+ +
  f^{\rm eq}_-)\right]
\nonumber\\
\left<k_0{\cal C}_0\right> 
&=& - \frac12\Gamma_m\left[(f_+ - f_-) - (f^{\rm eq}_+ - f^{\rm eq}_-)\right] \,,
\label{coll_terms_thermal}
\end{eqnarray}
where we have defined $f^{\rm eq}_\pm \equiv n_{\rm eq}(\pm \omega_{\vec k})$, and we have neglected terms of order ${\cal O}(\Gamma^2, \partial_t(m^2)\Gamma)$. The $\Gamma_i$-functions involve projections onto the mass- and the zero-momentum shells:
\begin{eqnarray}
\Gamma_m(|\vec{k}|,t) &\equiv& \Gamma(k_0=\omega_{\vec k}(t),|\vec{k}|)
\nonumber\\ 
\partial_{k_0}\Gamma_0(|\vec{k}|)
&\equiv& \partial_{k_0}\Gamma(k_0=0,|\vec{k}|)\,. 
\label{collfunctions}
\end{eqnarray}
Note in particular that the derivative term is computed at ``off-shell" value $k_0=0$ corresponding to the coherence shell.  Expressing $f$-functions in terms of the moment functions through eqs.~(\ref{rho-f_HOM}), and inserting the resulting expressions back to equations (\ref{rho_Eq_Coll1}) we eventually find a closed set of equations:
\begin{eqnarray}
\frac14 \partial_t^2 \rho_0 + \omega_{\vec k}^2 \rho_0 - \rho_2 &=& -\frac12\partial_{k_0}\Gamma_0\,\partial_t\rho_0
\nonumber\\
\partial_t\rho_1 &=& -\frac{1}{\omega_{\vec k}}\Gamma_m \left(\rho_1 - \rho_{1,{\rm
      eq}}\right)
\nonumber\\
\partial_t\rho_2 - \frac12 \partial_t(m^2) \rho_0 &=& -\frac{1}{\omega_{\vec k}}\Gamma_m \left(\rho_2 - \rho_{2,{\rm
      eq}}\right)\,.
\label{rho_Eq_Coll2}
\end{eqnarray}
eqs.~(\ref{rho_Eq_Coll2}) are the master equations that are used in
section \ref{sec:numexample} to study the coherent production of
unstable particles in an oscillatory background.

\section{Coherent production of unstable particles}
\label{sec:numexample}
In this section we shall compute the effect collisions on the coherent production of unstable scalar particles  in the presence of a time varying driving field. This problem is very similar to the one we considered for fermions in ref.~\cite{HKR2}. In fact we are taking the mass term driving the particle production to be of the same form as in~\cite{HKR2}:
\begin{equation}
m^2(t) \equiv |m_0 + e^{-\gamma \tau}\big(A \cos(2\omega_\varphi t) + i B\sin(2 \omega_\varphi t) \big)|^2\,,
\label{secondmassterm}
\end{equation}
where  $m_0$, $A$, $B$, $\omega_\varphi$ (oscillation frequency of the the driving field $\varphi$) and $\gamma$ are real constants. 
To illustrate more clearly the qualitative aspects of the method we take some parameters of the model in this example to be outside the adiabatic limit. Reader is warned that this may make quantitative results somewhat inaccurate. It should be emphasised that the method, especially the calculation of $n_{\vec k}$ is proven only in adiabatic limit.   The task is simply to compute the collision functions  $\Gamma_m$ and $\partial_{k_0}\Gamma_0$ appearing in (\ref{collfunctions}) for the particular model under consideration, and solve the equations (\ref{rho_Eq_Coll2}) with some suitable initial conditions. Here we shall consider the interaction
\begin{equation} 
{\cal L}_{\rm int} = - y\; \bar\psi \psi \phi\,, 
\label{interaction}
\end{equation}
where $\phi$ is the real scalar field whose dynamics we are interested in, and $\psi$ is some fermion field, which will be assumed to be in thermal equilibrium at all times. We assume that $m_\phi > 2 m_\psi$ at least for some $t$, and consider the effect of the induced instability on the $\phi$-particle production. The explicit expressions for $\Gamma_m$ and $\partial_{k_0}\Gamma_0$ that we will be using are computed in the appendix A.

Numerical solution of equations (\ref{rho_Eq_Coll2}) is not always straightforward; depending on the precise form of the driving term they can become very unstable against numerical errors. The problem can be traced to the third  equation in (\ref{KG_Eq_Coll1}). It turns out that for strong driving fields the dynamical evolution of $\rho_2$ is  very delicate and it is impossible to discern the physical solution from exponentially growing spurious numerical errors.  However, these  instabilities can be circumvented by transforming equations (\ref{rho_Eq_Coll2}) into an equivalent set of nonlinear equations in the presence of a constraint. Indeed, taking the sum of the first equation in (\ref{rho_Eq_Coll2}) multiplied by $ -2 \partial_t\rho_0$ and the third equation multiplied by $ 2\rho_0$ we will get first order differential equation: 
\begin{equation}
\partial_t X = -2 \partial_t\rho_0 \left<i{\cal C}_1\right> + 2 \rho_0 \left<k_0{\cal C}_0\right> \,,
\end{equation}
where
\begin{equation}
X = 2 \rho_2 \rho_0 - \omega_{\vec k}^2 \rho_0^2 - \frac14(\partial_t\rho_0)^2 \,.
\label{numerical_X}
\end{equation}
One observes that the function $X$ is a constant of motion in free theory, and so it should only change  slowly in the interacting theory providing collision integrals are small, which should be the case for the perturbation expansion to be valid. Initial value of $X$ can of course be calculated from initial values of $\rho_i$:s. The advantage of this formulation is that we can use the algebraic equation to solve $\rho_2$, while replacing the dynamical equation for $\rho_2$ by a 
much better behaving equation for $X$. That is, we use the equations
\begin{eqnarray}
\frac14 \partial_t Y + \omega_{\vec k}^2 \rho_0 - \rho_2  &=&  
 - \frac12\partial_{k_0}\Gamma_0 \, Y
\nonumber\\
\partial_t\rho_1 &=& 
- \frac{1}{\omega_{\vec k}}\Gamma_m \left(\rho_1 - \rho_{1,{\rm eq}}\right)  
\nonumber\\
\partial_t X &=& \partial_{k_0}\Gamma_0\,Y^2  
 - 2 \rho_0 \frac{\Gamma_m}{\omega_{\vec k}} \left( \rho_2 - \rho_{2,{\rm eq}}\right) \,,
\label{numerical_rho}
\end{eqnarray}
where $Y\equiv \partial_t \rho_0$ and $\rho_2 = (X+\sfrac14 Y^2+\omega_{\vec k}^2\rho_0^2)/2\rho_0$. Moreover, thermal values for $\rho_{1,2}$ can be seen from eqs.~(\ref{rho-f_HOM}) and (\ref{eqfspm})\footnote{The equation for $\rho_1$ in eqs.~(\ref{numerical_rho}) is actually trivial in the case of a real scalar field where $\rho_1=-\sfrac12$ throughout. However, these equations are valid as such also for a complex scalar field, where $\rho_1$ becomes a dynamical variable related to a chemical potential.}: $\rho_{1,\rm eq} = -\sfrac12$ and $\rho_{2,\rm eq} = \omega_{\vec k}( n_{\rm eq}(\omega_{\vec k})+\sfrac12)$. Equations (\ref{numerical_rho}) are numerically stable and easy to solve. In figure \ref{fig:numberdensity1} we show the evolution of the number density $n_{\vec k}$ and the absolute value of the coherence function $f_c(|\vec k|)$ as given by eq.~(\ref{f-rho_HOM}):
\begin{equation}
n_{\vec{k}} = \frac{1}{\omega_{\vec k}}\rho_2 - \frac{1}{2} 
\qquad  {\rm and} \qquad
f_c(|\vec{k}|) = \rho_0 - \frac{1}{\omega_{\vec k}^2}\rho_2 \,,
\end{equation}
under the influence of a driving mass term (\ref{secondmassterm}). The uppermost panel corresponds to case without collisions. The increase of the number density (thick solid line) is seen to be accompanied by a steady growth of the amplitude of the coherence (rapidly oscillating thin dotted blue line). In the middle panel we show the same solution in the case where we have included only the mass-shell collision terms, but set artificially $\partial_{k_0}\Gamma_0\equiv 0$. Here one sees that the number density decreases in the intervals between the stepwise growth as a result of decays. Finally the lowest panel shows the case where all collision terms are included properly. It is interesting that the growth of the number density is most strongly affected by the collisions acting on the coherence solution.  This can be understood when one solves for the time-evolution of $n_{\vec k}$ directly from (\ref{rho_Eq_Coll2}). The result is:
\FIGURE[t]{\includegraphics[width=12.5cm]{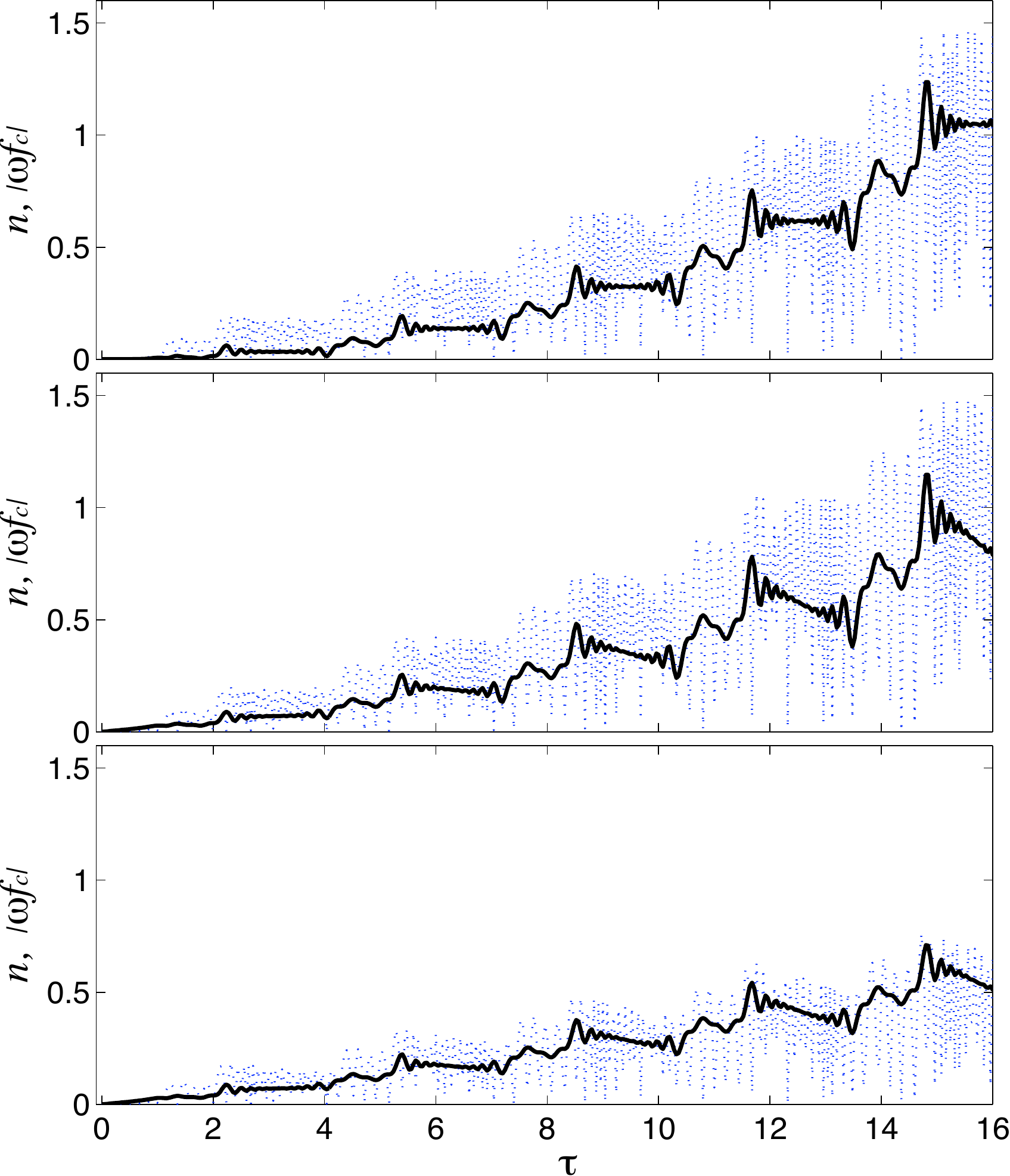}
        \caption{Shown is the number density $n_{{\vec k}}$ (thick solid line) and the coherence function $f_c(|\vec k|)$ (dotted(blue) line) with changing interactions. The driving mass function is taken to be $m^2(t) = |(1 + 1.5 \cos(2 \omega_{\varphi} t) + i\, 0.1 \sin(2 \omega_{\varphi} t))T|^2$. The upper panel corresponds to the case without collisions. In the central panel we have included the collision terms on the mass-shells, but kept $\partial_{k_0} \Gamma_0 = 0$. In the lowest panel the full interaction terms were kept for all shells. For parameters we have used $|\vec k| = 0.6\, T$, $y = 1$, $m_\psi=0.1\, T$ and $\omega_\varphi = 0.1 T$, where temperature $T$ sets the scale.  At $\tau \equiv \omega_\varphi t = 0$ the system is in the adiabatic vacuum.}
        \label{fig:numberdensity1}}
\begin{equation}
\partial_t n_{\vec k} = (\partial_t \omega_{\vec k}) f_c  -\frac{1}{\omega_{\vec k}}\Gamma_m( n_{\vec k} -  n_{\rm eq\, \vec k})\,.
\label{vemmu}
\end{equation}
That is, the creation rate of $n_{\vec k}$ is completely controlled by the coherence. Thus collisions that destroy the coherence also directly cut down 
the {\em growth} of the number density. This is potentially more important effect than the decay of the already created particle number effected by 
the on-shell decay rate $\Gamma_m$.

From equations (\ref{numerical_rho}) it is not obvious how collisions affect the coherence. However, it can be shown that their only stationary solution corresponds to $Y_{\rm eq}=0$ and $X_{\rm eq} = (n_{\rm eq}(\omega_{\vec k})+ \sfrac12)^2$, and this equilibrium limit can only be reached only through the effect of $\partial_{k_0} \Gamma_0$-terms. That this solution corresponds to vanishing coherence, becomes most evident when the evolution equations are written directly in terms of $f_c$. The general form of eqs.~(\ref{numerical_rho}) is quite messy when written in terms of the $f_i$-functions, but  they become much simpler if we assume that the driving mass term is a constant. In this limit the growth term for $n_{\vec k}$ vanishes in eq.(\ref{vemmu}) and the equation for $f_c$ becomes:
\begin{equation}
\partial_t^2 f_c + 4 \omega_{\vec k} f_c = - 2\partial_{k_0} \Gamma_0 \partial_t f_c \,.
\end{equation}
That is, when the driving term is shut off, the particle number decays towards the equilibrium value with a rate given by $\Gamma_m$ as expected. Simultaneously the coherence is driven to zero by the interaction term $\partial_{k_0} \Gamma_0$. Note that collisions act as a {\em friction} on coherence, just as in the case with fermions~\cite{HKR2}. Thus the solution is an oscillating function with an exponentially dying amplitude. We demonstrate this behavior from the full equations in figure \ref{fig:numberdensity2} for a particular initial configuration created from vacuum  by the same driving term used in figure \ref{fig:numberdensity1}.
\FIGURE[t]{\includegraphics[width=12.5cm]{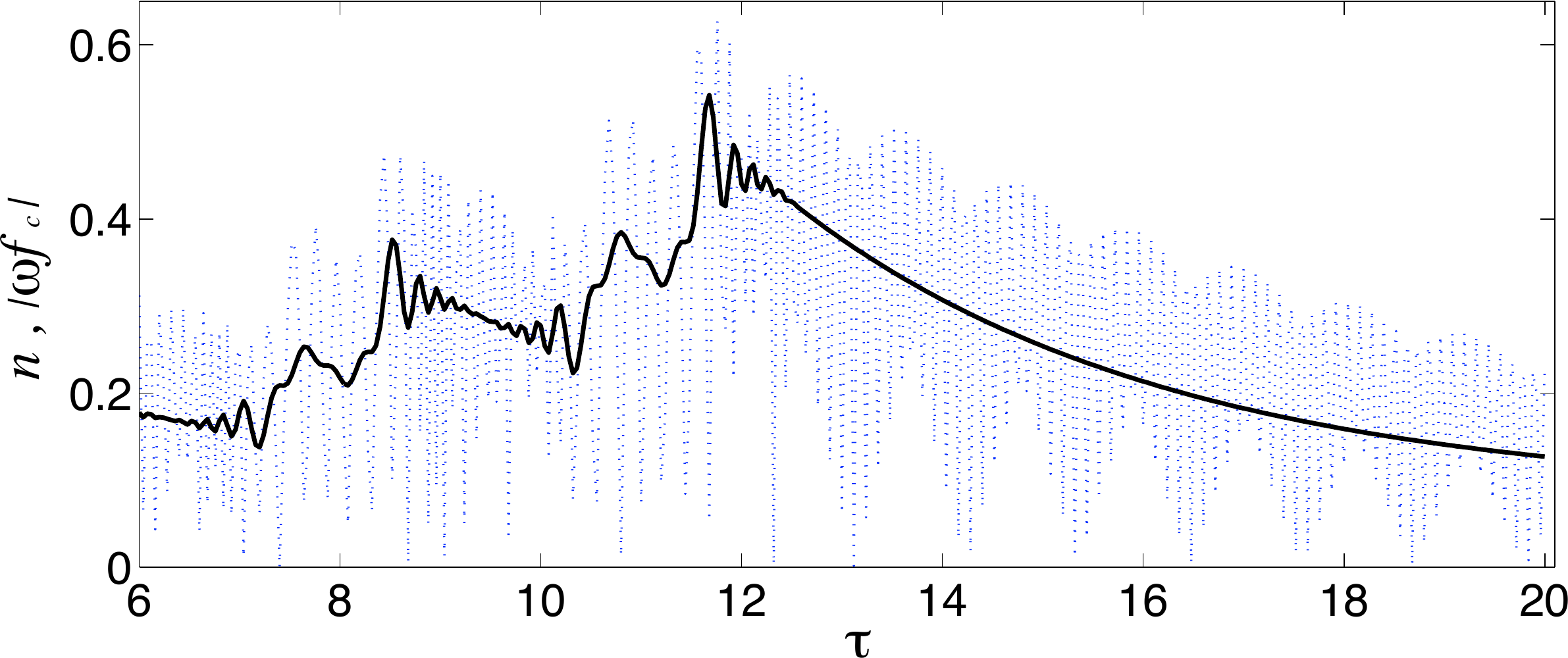}
        \caption{The same configuration as in the lowest panel of figure \ref{fig:numberdensity1} but with driving mass term smoothly set to constant $m = 2.5 \,T $ after $\tau > 4 \pi$.}
        \label{fig:numberdensity2}}

Finally, in figure \ref{fig:numberdensity3} we show the evolution of the number density under the influence of a driving force whose amplitude decays in time with the exponent $\gamma= 0.05$. The black solid line corresponds to a noninteracting case, and the thin solid (green), dash-dotted (blue) and dashed (red) lines to interacting cases with increasing strength of interaction. The dotted line corresponds to the local equilibrium number density $n_{\rm eq}$. This function oscillates as a result of the time-varying mass of the field. The steady growth of the particle number is now absent, as a result of the non-resonant nature of the driving field. The instantaneous particle number is mostly controlled by the quantum coherence effects when the interactions are absent or weak. In the case with weakest interactions the particle number is still strongly modulated by the background field, but there is a clear decaying trend due to interactions. In the case shown by dash-dotted (blue) curve the interactions are so strong that they almost eliminate the coherence after the first peak and the tendency of interactions to push the particle number towards their equilibrium values is beginning to show. This tendency is even clearer in for  the dashed (red) curve corresponding to most strongly interacting case. In the limit of infinitely strong interactions no coherence would be left and the particle number would follow the local adiabatic 
equilibrium particle number. 

\FIGURE[t]{\includegraphics[width=12.5cm]{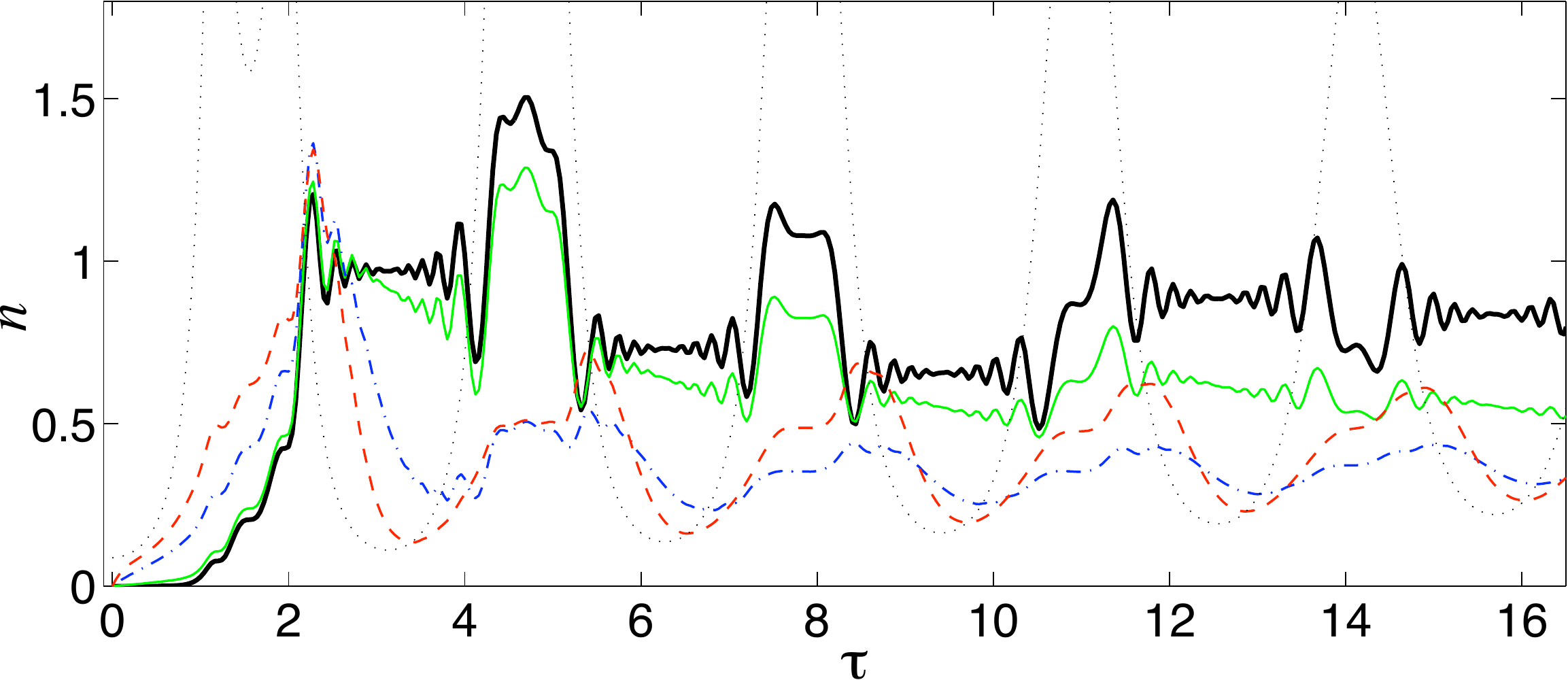}
        \caption{Number densities of particles in same configuration as before but with $|\vec k| = 0.3\, T$ and mass oscillation amplitude suppressed by exponent with $\gamma = 0.05$. The black solid line corresponds to a noninteracting case $ y = 0$, and the thin solid (green) $ y = 1$, dash-dotted (blue) $ y = 3$ and dashed (red) $ y = 5$ lines cases with different interaction strengths. The thin dotted line corresponds to the local equilibrium number density $n_{\rm eq}$.}
        \label{fig:numberdensity3}}

These examples show that using our methods it is possible to describe coherent particle production in the presence of decohering interactions. We considered only the case of decay, but it would be straightforward to extend the treatment to the case of collisions. The effect of collisions was qualitatively different on the mass-shell and on the coherence solutions, since in the latter case interactions introduce a friction term that tends to erase the oscillating coherence function, whereas the mass shell distributions were found to feel the usual relaxation towards equilibrium.

\section{Conclusions and outlook}
\label{sec:discussion}
In this paper we have derived quantum transport equations including nonlocal coherence effects for relativistic and nonrelativistic scalar fields in systems with particular spacetime symmetries. The time-dependent but homogeneous systems include for example particle production during phase transitions or during the inflation in the early universe. The static, planar symmetric problems include for example reflection off a potential, such as Klein problem, or off a mass wall induced by a phase transition front in the early universe.

The key observation leading to a calculable approximation scheme was the observation that in both of these geometries the 2-point correlator, written in the mixed representation, has new spectral solutions living on shell $k_0 = 0$ in the homogeneous case and on shell $k_z = 0$ in the static planar symmetric case. These solutions were interpreted to carry information on the nonlocal coherence between particle and antiparticle excitations in the former and between left- and right moving states in the latter case. We demonstrated the physical nature of these new spectral solutions by applying the formalism to exactly solvable models, such as relativistic Klein problem and bound states in one dimensional nonrelativistic potential well.  

The nontrivial singularity structures described above were found as an approximation to collisionless equations to the lowest order in gradients. The core of our calculational scheme was the argument that these structures should provide a reasonable {\em ansatz} for the 2-point function when relating the moments to physical on-shell functions at the boundary of the system and when computing collision terms in the moment expansion of the Kadanoff-Baym equation.  When the ansatz is introduced into the collision terms they become completely parametrized by the on-shell functions in the ansatz. Because the on-shell functions can be uniquely related to the lowest moment functions, the resulting moment equations  can be solved. Despite the simplicity of the resulting equations, they contain the essential information about the evolution of nonlocal quantum coherence under decohering interactions. Based on the nature of the approximation we argue that our method could be useful even when the background field is not necessarily slowly varying. Method requires the existence of an adiabatic boundary, or boundaries where the on-shell functions can be related to the moment functions. It is thus best suited to problems with localized disturbances in background field configurations with asymptotic adiabatic boundaries. 

Our method provides a natural definition for the adiabatic particle number $n_{\vec k}$ related to the value of the phase-space functions multiplying the singular mass-shells. This definition was applied to the particle number evolution during inflation where it was shown to correspond to the adiabatic particle number defined \eg in ref.~\cite{BirrelAndDavis}. Moreover, our particle number coincided with the definition of ref.~\cite{Berges} for slowly varying fields. We also computed the particle number evolution in the presence of a driving background interaction in the form of a time-dependent mass term. This situation could model for example the particle creation during phase transitions or at the end of the inflation.  We then included decoherence assuming that the scalar particles created by the background fields were unstable. We found that the effect of interactions on particle number divided to two parts: first the existing particle number was suppressed by decays as expected and secondly interactions provided a friction term on the growth of the coherence. However, as the growth of $n_{\vec k}$ was found to be completely controlled by coherence, the friction term turned out to be most efficient in reducing the particle number created by unit time.

Many of the results presented here were qualitatively similar to those derived earlier by us~\cite{HKR1,HKR2} for fermions. However, details of the derivation were substantially different so as to warrant a complete independent treatment in this paper. It would be interesting to apply our formalism to study for example the effect of collisions or decays on the production of scalar fields in a realistic model for a parametric resonance. It should also be straightforward to extend our nonrelativistic formulation for example for 3D-cubic lattice potentials and study atoms in such lattices under the influence of external thermal noise.

\section*{Acknowledgements}
This work was partly supported by a grant from the Jenny and Antti
Wihuri Foundation (Herranen) and the Magnus Ehrnrooth foundation
(Rahkila).

\appendix
\section{Yukawa interaction with thermal background}
\label{sec:self-energy}

In this appendix we compute the appropriate self-energies for the use
of the master equations (\ref{rho_Eq_Coll2}) in the case of a Yukawa interaction
with thermal background. We start with the interaction Lagrangian:
\begin{equation} 
{\cal L}_{\rm int} = - y\; \bar\psi \psi \phi 
\label{interactionA}
\end{equation}
where $\phi$ is the considered real scalar field and $\psi$ is some
fermion field. We use the 2PI effective action method
to calculate the self-energies (\ref{2PIsigmasC}) at 1-loop level. The
lowest order 2PI-graph based on interaction (\ref{interactionA})
(figure~\ref{fig:loop}) gives the contribution 
\begin{equation} 
\Gamma_{\rm 2PI} = -y^2 \int_C d^4u\,d^4v\, {\rm Tr}\left[G(u,v)
  G(v,u)\right]\prop(u,v) \,, 
\label{gamma2pI}
\end{equation}
where $G$ is the propagator of the fermion $\psi$ and the integration
is over the Keldysh time path. From this we get the self-energies by use of
eq.~(\ref{2PIsigmas}). In particular, after performing the Wigner
transformations we have:
\begin{equation} 
  i\Pi^{<,>}(k,x) = - y^2 \int \frac{d^4k'}{(2\pi)^4} {\rm
    Tr}\left[G^{>,<}(k',x) G^{<,>}(k+k',x)\right]\,.
\end{equation}
We assume thermal background so that the fermion $\psi$ distributions
appearing in the loop are thermal. The appropriate thermal propagators with
real constant mass are (see for example \cite{PSW}):
\begin{eqnarray}
  iG_{\rm eq}^<(k) &=& 
   2\pi\,{\rm sgn}(k_0)\left( \kdag + m_f \right) 
        n^f_{\rm eq}(k_0)\delta(k^2-m_f^2)           
\nonumber \\ 
 iG_{\rm eq}^>(k) &=& 
   2\pi\,{\rm sgn}(k_0)\left( \kdag + m_f \right) 
         (1 - n^f_{\rm eq}(k_0))\delta(k^2-m_f^2) \,,
\label{thermal_prop_fermion}
\end{eqnarray}
where $n^f_{\rm eq}(k_0) \equiv 1 / (e^{\beta k_0} + 1)$ is the
standard Fermi-Dirac distribution function. 

In our present analysis we need to evaluate the
self-energies $\Pi^{<,>}$ both on the mass-shell $k_0^2-\vec{k}^2=m(t)^2$
as well as on the $k_0=0$-shell. On the mass-shell we get:
\begin{equation} 
i\Pi^<(k_0=\pm\omega_{\vec k}(t),|\vec{k}|) = \frac{|y|^2 T}{4\pi
  |\vec{k}|}\frac{\lambda_m}{m^2}\,\theta(\lambda_m)
   \int_{\alpha - \beta}^{\alpha + \beta} {\rm d}y \: 
       \frac{1}{(e^y + 1) (e^{k_0/T - y} + 1 )}  \,, 
\label{Pi_ms}
\end{equation}    
with
\begin{equation}
\alpha = \frac{k_0}{2T} \qquad\quad {\rm and} \qquad
\beta  = \frac{\lambda_m^{1/2}}{2m^2}
\frac{|\vec{k}|}{T}\,,
\end{equation}
\FIGURE[t]{\includegraphics[width=8.8cm]{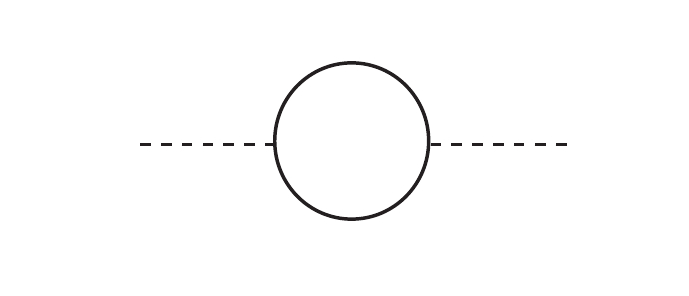}
        \caption{The only diagram contributing to the self-energy at 1-loop level.}
        \label{fig:loop}}
where $\lambda_m \equiv \lambda(m^2,m_f^2,m_f^2) \equiv m^2(m^2 - 4
m_f^2)$ is the usual kinematic phase space function on the mass-shell. On the
$k_0=0$-shell we get instead: 
\begin{equation} 
i\Pi^<(k_0=0,|\vec{k}|) = \frac{|y|^2
  T}{4\pi|\vec{k}|}\frac{\lambda_0}{\vec{k}^2}\bigg(1-\tanh\Big(\frac{\lambda_0^{1/2}}{4|\vec{k}|T}\Big)\bigg)\,,
\label{Pi_k0}
\end{equation}   
where now $\lambda_0 \equiv \lambda(-\vec{k}^2,m_f^2,m_f^2) \equiv
\vec{k}^2(\vec{k}^2 + 4 m_f^2)$.

Since we are computing $\Pi^{ab}$'s in the thermal limit, the expression 
for $\Pi^>$ can be obtained from that for $\Pi^<$ by use of the 
Kubo-Martin-Schwinger (KMS) relation:
\begin{equation} 
\Pi^>(k) = e^{\beta k_0} \Pi^<(k) \,, 
\label{KMSrelaatio}
\end{equation} 
This relation follows directly from the corresponding relation between the
thermal equilibrium propagators $G_{\rm eq}^{<,>}$
eq.~(\ref{thermal_prop_fermion}). Using this relation we find that
\begin{equation}
\Gamma(k) = \frac{i}{2}(1+e^{\beta k_0} ) \Pi^<(k) \,.
\label{gammakms}
\end{equation}
Now a direct computation shows that
\begin{equation}
\Gamma(-k_0,|\vec{k}|) = -\Gamma(k_0,|\vec{k}|) \qquad {\rm and}
\qquad \Gamma(k_0=0,|\vec{k}|) = 0\,,
\label{gamma_relaatio1}
\end{equation}
such that the assumptions made in section \ref{sec:interactions} are
indeed verified for this type of interaction. For the use of the
master equations (\ref{rho_Eq_Coll2}) we still need the
derivative $\partial_{k_0}\Gamma_0(|\vec{k}|)
\equiv \partial_{k_0}\Gamma(k_0=0,|\vec{k}|)$. A direct computation
gives
\begin{equation}
\partial_{k_0}\Gamma(k_0=0,|\vec{k}|) = \frac{1}{2T}i\Pi^<(k_0=0,|\vec{k}|)\,.
\label{gamma_relaatio2}
\end{equation}
Expressions (\ref{Pi_ms}) and (\ref{Pi_k0}) together with the
relations (\ref{gammakms})-(\ref{gamma_relaatio2}) complete the
computation of all required self-energy functions for the
use of the master equations (\ref{rho_Eq_Coll2}).

%
%

\end{document}